\documentclass[twocolumn]{aastex63}
\received{\today}
\revised{\today}
\accepted{XXX}
\submitjournal{ApJL}

\shorttitle{Quiescent galaxy in a protocluster at $z=3.09$}
\shortauthors{Kubo et al.}

\begin{document}

\title{A massive quiescent galaxy confirmed in a protocluster at $\bf z=3.09$}

\correspondingauthor{Mariko Kubo}
\email{kubo@cosmos.phys.sci.ehime-u.ac.jp}

\author[0000-0000-0000-0000]{Mariko Kubo}
\affiliation{Ehime University, 2-5 Bunkyo-cho, Matsuyama, Ehime, Japan}
\affiliation{National Astronomical Observatory of Japan, 2-21-1 Osawa, Mitaka, Tokyo 181-8588, Japan}
\author[0000-0000-0000-0000]{Hideki Umehata}
\affiliation{RIKEN Cluster for Pioneering Research, 2-1 Hirosawa, Wako-shi, Saitama 351-0198, Japan}
\affiliation{Institute of Astronomy, School of Science, The University of Tokyo, 2-21-1 Osawa, Mitaka, Tokyo 181-0015, Japan}
\author[0000-0000-0000-0000]{Yuichi Matsuda}
\affiliation{National Astronomical Observatory of Japan, 2-21-1 Osawa, Mitaka, Tokyo 181-8588, Japan}
\affiliation{Department of Astronomy, School of Science, The Graduate University for Advanced Studies, SOKENDAI, Mitaka, Tokyo 181-8588, Japan}
\author[0000-0000-0000-0000]{Masaru Kajisawa}
\affiliation{Ehime University, 2-5 Bunkyo-cho, Matsuyama, Ehime, Japan}
\author[0000-0000-0000-0000]{Charles C. Steidel}
\affiliation{Cahill Center for Astronomy and Astrophysics, California Institute of Technology, MC 249-17, Pasadena, CA 91125, USA}
\author[0000-0000-0000-0000]{Toru Yamada}
\affiliation{Institute of Space an Aeronautical Science, Japanese Aerospace Exploration Agency, 3-1-1, Yoshinodai, Chuo-ku, Sagamihara, Kanagawa, 252-5210, Japan}
\author[0000-0000-0000-0000]{Ichi Tanaka}
\affiliation{Subaru Telescope, National Astronomical Observatory of Japan, 650 North A’ohoku Place, Hilo, Hawaii, 96720, U.S.A.}
\author[0000-0000-0000-0000]{Bunyo Hatsukade}
\affiliation{Institute of Astronomy, School of Science, The University of Tokyo, 2-21-1 Osawa, Mitaka, Tokyo 181-0015, Japan}
\author[0000-0000-0000-0000]{Yoichi Tamura}
\affiliation{Department of Physics, Nagoya University, Furo-cho, Chikusa-ku, Nagoya, Aichi 464-8601, Japan}
\author[0000-0000-0000-0000]{Kouichiro Nakanishi}
\affiliation{National Astronomical Observatory of Japan, 2-21-1 Osawa, Mitaka, Tokyo 181-8588, Japan}
\affiliation{Department of Astronomy, School of Science, The Graduate University for Advanced Studies, SOKENDAI, Mitaka, Tokyo 181-8588, Japan}
\author[0000-0000-0000-0000]{Kotaro Kohno}
\affiliation{Institute of Astronomy, School of Science, The University of Tokyo, 2-21-1 Osawa, Mitaka, Tokyo 181-0015, Japan}
\affiliation{Research Center for the Early Universe, The University of Tokyo, 7-3-1 Hongo, Bunkyo-ku, Tokyo 113-0033,Japan}
\author[0000-0000-0000-0000]{Chien-Feng LEE}
\affiliation{Institute of Astronomy, School of Science, The University of Tokyo, 2-21-1 Osawa, Mitaka, Tokyo 181-0015, Japan}
\author[0000-0000-0000-0000]{Keiichi Matsuda}
\affiliation{Department of Physics, Nagoya University, Furo-cho, Chikusa-ku, Nagoya, Aichi 464-8601, Japan}

\nocollaboration{13}



\begin{abstract}

We report a massive quiescent galaxy at $z_{\rm spec}=3.0922^{+0.008}_{-0.004}$ 
spectroscopically confirmed at a protocluster in the SSA22 field 
by detecting the Balmer and Ca {\footnotesize II} absorption features with multi-object 
spectrometer for infrared exploration (MOSFIRE) on the Keck I  telescope. 
This is the most distant quiescent galaxy confirmed in a protocluster to date. 
We fit the optical to mid-infrared photometry and spectrum simultaneously 
with spectral energy distribution (SED) models of parametric and
nonparametric star formation histories (SFH).
Both models fit the observed SED well and confirm that this object is a massive quiescent galaxy 
with the stellar mass of $\log(\rm M_{\star}/M_{\odot}) = 11.26^{+0.03}_{-0.04}$ and $11.54^{+0.03}_{-0.00}$, 
and star formation rate of $\rm SFR/M_{\odot}~yr^{-1} <0.3$ and $=0.01^{+0.03}_{-0.01}$ 
for parametric and nonparametric models, respectively. 
The SFH from the former modeling is described as an instantaneous starburst 
while that of the latter modeling is longer-lived but  
both models agree with a sudden quenching of the star formation at $\sim0.6$ Gyr ago. 
This massive quiescent galaxy is confirmed in an extremely dense group of galaxies 
predicted as a progenitor of a brightest cluster galaxy formed via multiple mergers in cosmological numerical simulations. 
We newly find three plausible [O {\footnotesize III}]$\lambda$5007 
emitters at $3.0791\leq z_{\rm spec}\leq3.0833$ happened to be detected around the target. 
Two of them just between the target and its nearest massive galaxy are possible evidence of their interactions. 
They suggest the future strong size and stellar mass evolution of this massive quiescent galaxy via mergers. 

\end{abstract}


\section{Introduction} \label{sec:intro}

The deep multi-wavelength imaging surveys have uncovered 
wide variety in galaxy population at high redshift. 
Lately, massive quiescent galaxies 
at up to $z\sim4$ have been discovered and confirmed spectroscopically
(e.g., \citealt{2005ApJ...626..680D, 2008ApJ...677L...5V,2009ApJ...700..221K, 2018A&A...618A..85S, 2019ApJ...885L..34T, 2020ApJ...889...93V,2020ApJ...903...47F}).
They are plausible progenitors of giant ellipticals today 
though their formation mechanism is a challenging problem 
since they have to form a stellar mass $M_{\star} \gtrsim 10^{11}~M_{\odot}$ 
and quench star formation in the early universe. 
Massive quiescent galaxies at high redshift are remarkably compact (a few to ten times smaller than local giant ellipticals) in general 
(e.g., \citealt{2005ApJ...626..680D, 2008ApJ...677L...5V,2009ApJ...700..221K, 2014ApJ...788...28V, 2018ApJ...867....1K,2021MNRAS.501.2659L}). 
To form them, the star formation accompanied with centrally concentrated intense starburst 
like gas rich major merger and violent disc instability is required 
(e.g., \citealt{2009ApJ...703..785D,2010ApJ...725.2324B,2014MNRAS.438.1870D,2015MNRAS.450.2327Z}). 
Lately, submillimeter galaxies (SMGs) similar in sizes, masses, and velocity dispersion 
with compact massive quiescent galaxies have been discovered 
and support this picture (e.g., \citealt{2014ApJ...782...68T,2016ApJ...832...19S,2020ApJ...889...93V}).
It should also be considered that galaxies assembled in the early universe 
or when the universe was much denser tend to be compact 
(e.g., \citealt{2014ApJ...780....1W,2015MNRAS.449..361W,2016MNRAS.456.1030W,2021ApJ...907L...8A}). 
Such massive quiescent galaxies at high redshift need strong size evolutions where  
evolutions of mass and size via mergers
(e.g., \citealt{2009ApJ...697.1290B,2009ApJ...699L.178N}), 
expansion due to mass-loss driven by feedback (e.g., \citealt{2008ApJ...689L.101F,2010ApJ...718.1460F}), 
and/or evolution of stellar population \citep{2014ApJ...791...45V} scenarios have been proposed but not yet been constrained exactly.

Given the environmental dependance of galaxies today, 
we need to investigate the formation and evolution scenarios together with environments. 
However, massive quiescent galaxies at $z>2$ in previous studies 
are mostly found in general fields or their environments are poorly explored, 
though several studies reported possible overdensities around them
\citep{2015A&A...576L...6S,2017ApJ...841L...6B,2018A&A...611A..22S,2020arXiv201113700S}. 
To track the formation history of giant ellipticals generally in clusters of galaxies properly, 
we also need to study their progenitors in protoclusters. 
The appearance of red sequence in protoclusters at  $z\lesssim 3$ 
have been found by deep near-infrared (NIR) observations 
(e.g., \citealt{2007MNRAS.377.1717K,2008ApJ...680..224Z,2010A&A...509A..83D,2013ApJ...778..170K,2016ApJ...833L..20S,2020ApJ...899...79S}), 
and massive quiescent galaxies have been confirmed spectroscopically by 
detecting Balmer absorption features in clusters at up to $z\approx2$ \citep{2013ApJ...776....9G}. 
The environmental difference of galaxy population at $z\sim4$ is shown statistically 
by using the deep and wide optical survey of the Hyper Suprime-Cam Subaru Strategic 
Program survey \citep{2018PASJ...70S..12T,2019ApJ...887..214K,2020ApJ...899....5I}. 
However morphologies of only a small number of massive quiescent galaxies
have been studied (\citealt{2008ApJ...680..224Z, 2017MNRAS.469.2235K},
and also \citealt{2013ApJ...772..118S} and \citealt{2015ApJ...804..117M} including mature clusters at $z\sim2$),
and the detailed spectral characteristics of massive quiescent galaxies 
have not yet been characterized in protoclusters. 

We have studied a protocluster at $z=3.09$ in the SSA22 field \citep{1998ApJ...492..428S} 
which is known as one of the most significant structures at high redshift.
This structure was first identified by the overdensity of optically selected galaxies 
\citep{1998ApJ...492..428S,2004AJ....128.2073H,2005ApJ...634L.125M,2012AJ....143...79Y}
and in later, further characterized with the overdensity 
of active galactic nuclei (AGNs) selected in $X$-ray \citep{2009ApJ...691..687L, 2009MNRAS.400..299L}, 
SMGs \citep{2009Natur.459...61T,2014MNRAS.440.3462U,2015ApJ...815L...8U,2017ApJ...835...98U,2018PASJ...70...65U}, 
and dusty starburst and passively evolving galaxies selected photometrically 
based on the deep NIR imaging \citep{2013ApJ...778..170K}.
Thus this protocluster is an ideal laboratory used to witness 
the transition of starburst galaxies into quiescent galaxies. 
Particularly, the $2'$ x $3'$ region containing the brightest SMG of this field 
was deeply observed with Atacama large millimeter/submillimeter array (ALMA) 
and a significant overdensity of SMGs was discovered 
(\citealt{2015ApJ...815L...8U,2017ApJ...835...98U,2018PASJ...70...65U}; hereafter ADF22). 
They are clustering along the large scale ($\sim1$ Mpc) Ly$\alpha$ filament 
which indicates the supply of cold gas through cosmic web \citep{2019Sci...366...97U}. 
This field is plausibly the central region of the SSA22 protocluster 
also supported from the large scale redshift distribution of galaxies \citep{2005ApJ...634L.125M,2015ApJ...799...38K}. 

In this paper, we report the confirmation of 
a massive quiescent galaxy at the ADF22 field of the SSA22 protocluster 
by detecting the absorption features spectroscopically 
with multi-object spectrometer for infrared exploration 
(MOSFIRE;  \citealt{2012SPIE.8446E..0JM}) on the Keck I telescope. 
This paper is organized as follows: 
In Section 2, we describe the target and observation with MOSFIRE and data
analysis. In Section 3, we present the spectral energy distribution (SED) 
fittings and discoveries of new [O {\footnotesize III}] $\lambda5007$
emitters around this massive quiescent galaxy. 
In Section 4, we discuss the star formation history and future evolution scenario.
Section 5 is the conclusion. 
In this study, we adopt cosmological parameters $\Omega_{\rm m} =0.3$, 
$\Omega_{\rm \Lambda} =0.7$ and $H_0 =70$ km s$^{-1}$ Mpc$^{-1}$.  
We assume the \citet{2003PASP..115..763C} Initial Mass Function (IMF). 
Magnitudes are expressed in the AB system. 

\section{Data and analysis} \label{sec:style}

\subsection{Target}

Previously, we conducted deep and wide NIR imaging observations 
of the SSA22 protocluster 
with multi-object infrared camera and spectrograph 
(MOIRCS; \citealt{2006SPIE.6269E..16I,2008PASJ...60.1347S}) 
on the Subaru telescope
over a field of  $\approx112$ arcmin$^2$ for $K_s\approx24$ at 5$\sigma$ level 
and found an overdensity of massive galaxies based on the photometric redshifts ($z_{\rm phot}$)
estimated with the SED fitting of the optical to MIR photometry \citep{2013ApJ...778..170K}.
We further confirmed many of these candidates as protocluster members spectroscopically \citep{2015ApJ...799...38K}.
Among them, we selected candidate passively evolving galaxies 
based on the $i-K$ vs. $K-[4.5]$ color following \citet{2005ApJ...624L..81L}. 
These colors are similar to rest-frame $UVJ$ often used to select quiescent galaxies (e.g., \citealt{2013ApJ...770L..39W}). 
We selected galaxies satisfying $i-K>3$ and $K-[4.5]<0.5$, and  $2.6<z_{\rm phot}<3.6$ as candidate quiescent galaxies. 
The color criterion in \citet{2013ApJ...778..170K} was set to avoid the contamination of dusty starburst galaxies detected in 24 $\mu$m 
which distribute at the border of general color criterion for quiescent galaxies.

\begin{figure}
\gridline{\fig{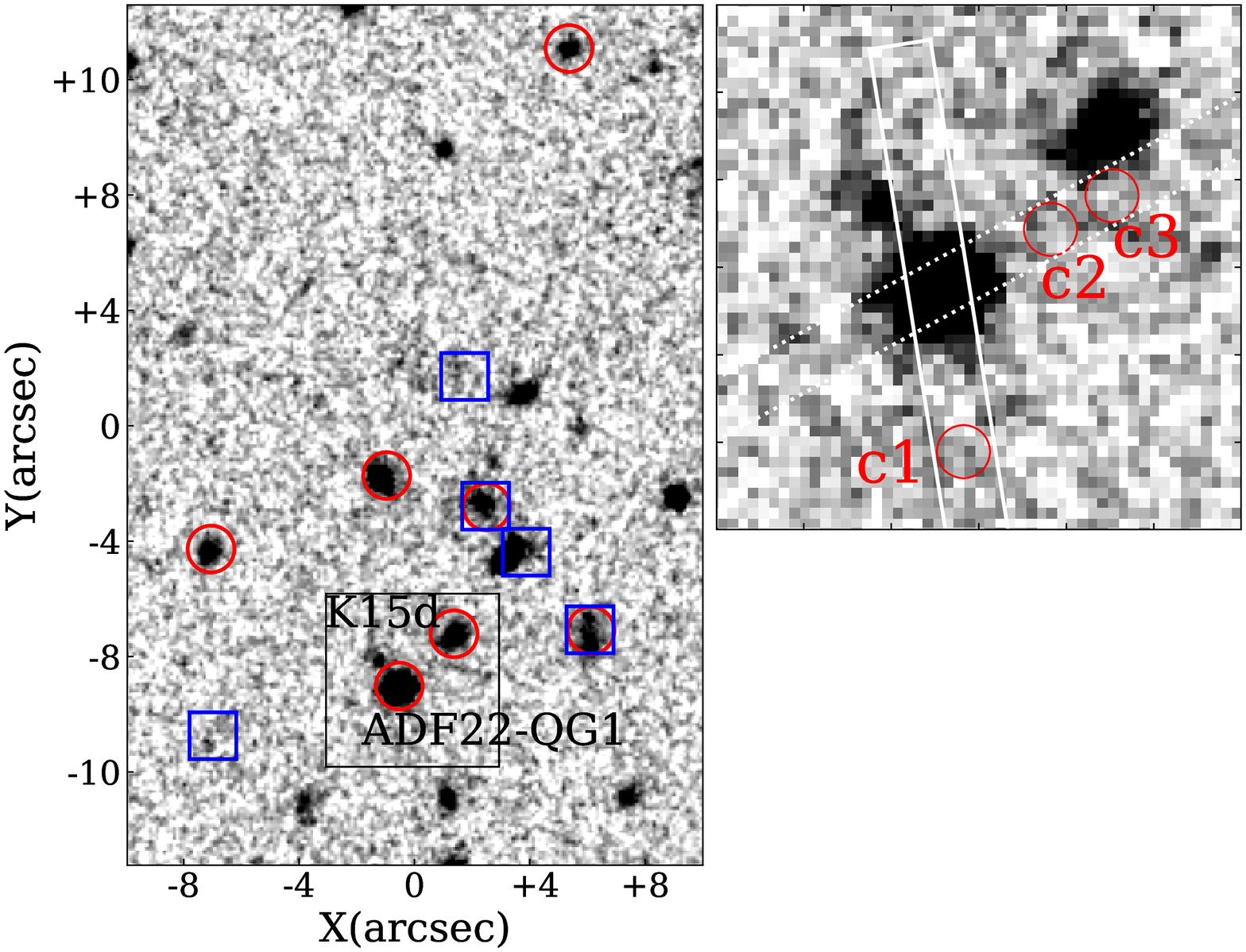}{0.5\textwidth}{}}
\caption{ {\it Left:} The MOIRCS $K_s$-band image of the target field (20.0 arcsec x 30.0 arcsec). 
This is the newly obtained image described in Section 2.3.
The red circles and blue squares show the objects spectroscopically 
confirmed at $3.0774\leq z_{\rm spec}\leq 3.0926$ with [O {\footnotesize III}] or Ly$\alpha$
(summarized in \citealt{2016MNRAS.455.3333K}), 
and with CO(3-2) \citep{2019Sci...366...97U} emission lines, respectively. 
{\it Right:} The zoom in image of the black squared region in the {\it left} panel 
which includes the ADF22-QG1 and adjacent massive galaxy, K15d (6.0 arcsec by side).
The white dashed and solid rectangles show the slit positions 
of $K$-17 and $K$($H$)-20 runs, respectively. 
The red circles (0.6 arcsec diameter) show the positions 
of plausible [O {\footnotesize III}]$\lambda$5007 
emitters detected simultaneously with ADF22-QG1 (Section 3.3). 
\label{fig:image}}
\end{figure}

Our target (R.A., Dec = 22:17:37.25, +00:18:16.0, hereafter ADF22-QG1) 
satisfies these selection criteria. 
It has a total magnitude of $K_{s,tot}=21.55$.  
The SED is well fitted with that of a quiescent galaxy namely 
single burst or exponentially declining star formation history (SFH) 
where star formation rate (SFR) $\propto \exp (-t/{\tau})$ with $\tau\sim0.1$ Gyr, 
and with an age of $\gtrsim 1$ Gyr at $z\sim3.1$.
We show the target on a rest-frame $UVJ$ color diagram in Fig. \ref{fig:uvj}.
Its spectroscopic redshift is limited to $z_{\rm spec} = 2.7$ 
or $3.0-3.15$ by detecting the 4000 \AA~break  \citep{2015ApJ...799...38K}.
We took a high-resolution $K'$-band image of the target by using the adaptive optics AO188 and 
infrared camera and spectrograph (IRCS) on the Subaru telescope and found that its
effective radius ($r_{eff}$) is $1.01\pm0.04$ kpc \citep{2017MNRAS.469.2235K}.
Uniquely, it is likely in an extremely dense group of massive galaxies and SMGs 
identified by follow-up NIR spectroscopic observations of a 1.1 mm source found with ASTE/AzTEC
(\citealt{2016MNRAS.455.3333K}; here after AzTEC14 group). 
The {\it left} panel of Fig. \ref{fig:image} shows the MOIRCS $K_s$-band image 
(described in \S~2.3) for 20 arcsec x 30 arcsec around the target. 
Except for our target, there are nine galaxies at $3.0774\leq z_{\rm spec}\leq 3.0926$
confirmed by detecting the [O {\footnotesize III}] $\lambda5007$ and/or CO(3-2) emission lines
\citep{2016MNRAS.455.3333K,2018PASJ...70...65U}.
Assuming that they are hosted by the same halo, 
its expected halo mass is $\gtrsim10^{13}~M_{\odot}$ based on the velocity distribution \citep{2016MNRAS.455.3333K}.

\subsection{Observation with MOSFIRE}

We observed the target with MOSFIRE in $K$-band in September 2017 and 2020, 
and in $H$-band in September 2020  (Hereafter $K$-17, $H$-20 and $K$-20).
The {\it right} panel of Fig. \ref{fig:image} shows the slit positions.
The observations and data reduction are described 
in \citet{2019Sci...366...97U} ($K$-17) and in H. Umehata et al. in prep ($H$-20 and $K$-20).
Briefly, the seeing conditions were $0.7\sim0.9$, $0.6$ and $0.6$ 
arcsec for $K$-17, $K$-20 and $H$-20 runs. 
We adopted the 2-position mask-nod sequence with $1.5$ arcsec dithers along slits 
with 120 (180) sec for each exposure in $H(K)$-band. 
The total exposure times at ADF22-QG1 were 2.5 and 6.5 h (3.2 h for $K$-17 
and 3.3 h for $K$-20) in $H$ and $K$-band, respectively.
The spectra were reduced with the publicly available MOSFIRE 
data reduction pipeline, {\sf MCSDRP} \citep{2014ApJ...795..165S}. 

We extracted the one-dimensional spectrum as follows 
to measure the fluxes of the spectrum and photometry consistently.
Here we calibrated the spectra to the $H$ and $K_s$-band 
photometry taken with MOIRCS (Section 2.4). 
First, $H$-band spectrum was re-binned to match the pixel scale 
with that in $K$-band ($\approx 0''.18$/pix and 2.17~\AA/pix) 
by using the {\sf magnify} task of {\sf IRAF}.
Next, we combined the target and error spectra for $\approx2.0$ arcsec in the spatial direction 
by an inverse variance weighting. 
Then, we measured the $H$ and $K_s$-band fluxes of the spectra 
by applying the transmission curve of MOIRCS $H$ and $K_s$-band filters.
Finally, we corrected the spectra to match these fluxes with the photometric fluxes. 

\subsection{Deeper K$_s$-band image taken with MOIRCS}

We newly obtained a deeper $K_s$-band image 
with MOIRCS which was upgraded in 2015 
(nuMOIRCS;  \citealt{2016SPIE.9908E..2GW,2016SPIE.9908E..28F}) on the Subaru telescope in July 2020 (PI: H. Umehata).
We collected the images with PSF FWHM of $0.3-0.6$ arcsec for 6.6 h total exposure time.
The data is reduced with the MOIRCS data reduction pipeline 
MCSRED\footnote{https://www.naoj.org/staff/ichi/MCSRED/mcsred\_e.html}.
We reduce the data in a standard procedure 
but the flux is calibrated to our previous $K_s$-band image 
taken with MOIRCS \citep{2013ApJ...778..170K} 
which is calibrated to UKIRT Infrared Deep Sky Survey (UKIDSS).
The zero-point error is $\approx0.05$ mag relative to UKIDSS. 
Briefly, the final combined image has a PSF FWHM of 0.40 arcsec 
and the $5 \sigma$ limiting magnitude at 0.80 arcsec diameter of 25.34 mag. 

\begin{figure*}
\gridline{\fig{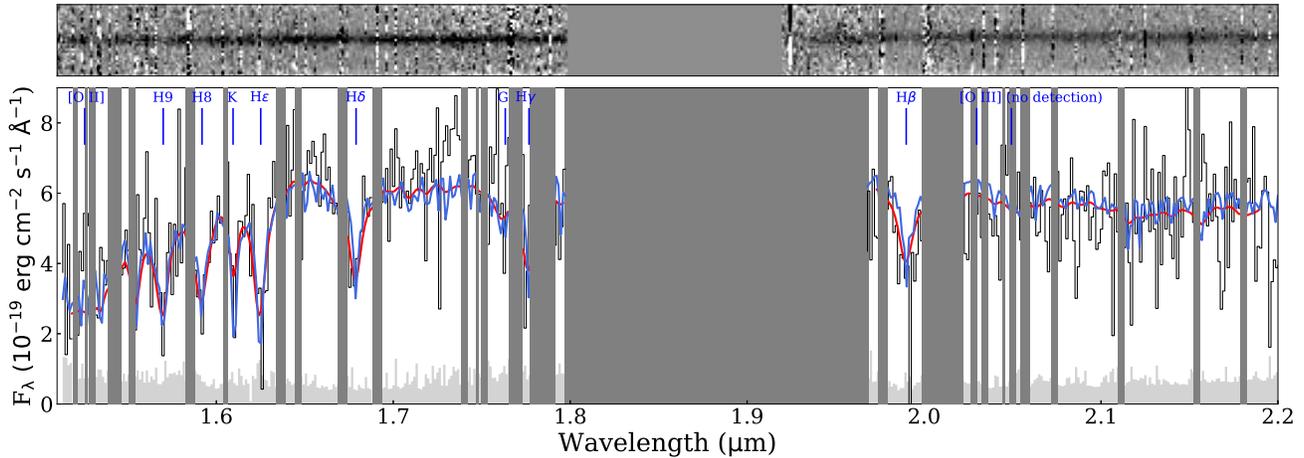}{1.0\textwidth}{}}
\caption{{\it Top:} The spectrum of ADF22-QG1 in the $H$ and $K$-band re-binned with $\approx13$ \AA~(6 pixels). 
{\it Bottom:} The black step plot shows the one-dimensional spectrum. 
The light gray filled histogram shows the 1$\sigma$ noise at each bin. 
The red and blue curves show the best-fit SEDs obtained with {\sf FAST++}
and continuity SFH prior of {\sf Prospector}, respectively.
The location of the notable absorption features, 
and [O {\footnotesize II}] and [O {\footnotesize III}] (no detection) emission lines is labeled. 
\label{fig:spectra}}
\end{figure*}

\begin{figure}
\gridline{\fig{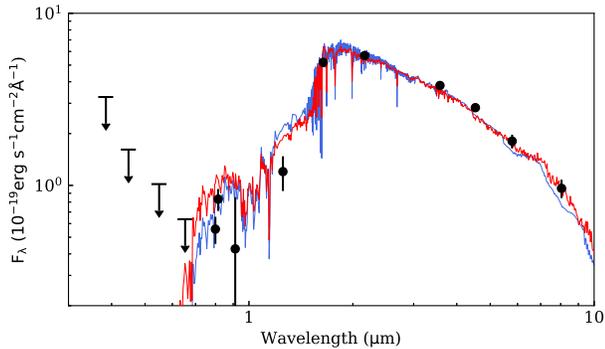}{0.5\textwidth}{}}
\caption{The black filled circles show the observed photometry of ADF22-QG1 with 1 $\sigma$ errors. 
The red and blue curves show the best-fit SEDs obtained with {\sf FAST++} 
and continuity SFH prior of {\sf Prospector}, respectively.
\label{fig:phot}}
\end{figure}

\subsection{Spectroscopic redshift} \label{subsec:redshift}

Figure \ref{fig:spectra} shows the spectrum of ADF22-QG1. 
There are clear Balmer and  Ca {\footnotesize II} absorption features, 
and an [O {\footnotesize II}]$\lambda$3727 emission line 
while [O {\footnotesize III}]$\lambda\lambda$4959, 5007 emission lines are not detected significantly.
Following \citet{2018A&A...618A..85S},  
the spectroscopic redshift is estimated with {\sf SLINEFIT}\footnote{https://github.com/cschreib/slinefit}  
which fits a spectrum with combinations of a stellar continuum template and emission/absorption lines
to find a redshift and line properties by a $\chi^2$ minimization procedure. 
It returns a spectroscopic redshift $z_{\rm spec}=3.0922^{+0.0008}_{-0.0004}$ which is consistent with the absorption features. 
Thus ADF22-QG1 is certainly in the SSA22 protocluster, 
moreover a member of the dense group of galaxies discussed in Section 4.2. 
This is the most distant quiescent galaxy detected of Balmer absorption features in a protocluster. 
Hereafter we adopt this spectroscopic redshift for ADF22-QG1. 

The [O {\footnotesize II}] flux of  ADF22-QG1 is $F_{\rm [OII]} = 5.7\pm1.0\times 10^{-18}$ erg s$^{-1}$ cm$^{-2}$.
The SFR from [O {\footnotesize II}] luminosity is calculated as follows; 
First, the SFR from H$\beta$ luminosity is computed 
assuming the case B recombination value intrinsic H$\alpha$/H$\beta$ = 2.86 \citep{2006agna.book.....O}, 
and the SFR to H$\alpha$ luminosity relation in \citet{2012ARA&A..50..531K},

$$\rm \log SFR_{\rm H\alpha}(M_{\odot}~\rm yr^{-1})=\log L(\rm H\alpha)(erg~s^{-1})-41.27. $$

The normal star forming galaxies (SFGs) at $z\sim3$ have 
log [O{\footnotesize II}]/H$\beta =0.0-0.6$ at most according to \citet{2016ApJ...822...42O}.
Then the SFR from [O{\footnotesize II}] is computed as,

$$\rm SFR_{\rm [O~II]}(M_{\odot}~yr^{-1})=4-15\times10^{-42}~L_{\rm [OII]}(erg~s^{-1}).$$ 

The SFR$_{\rm [O~II]}$ for ADF22-QG1 is $2-7~\rm M_{\odot}~\rm yr^{-1}$ for $A_V=0$ 
and $4-14~\rm M_{\odot}~\rm yr^{-1}$ for $A_V=0.5$ 
which are more than ten times higher than that from the SED fitting described in later.
[O {\footnotesize II}] emission lines are frequently seen in quiescent galaxies at high redshift 
and the SFRs from [O {\footnotesize II}] are tend to be higher than 
those from SED modeling 
(e.g., \citealt{2006ApJ...648..281Y,2010ApJ...716..970L,2015ApJ...799..206B,2018A&A...618A..85S}). 
The [O {\footnotesize II}] flux of ADF22-QG1 is similar to those in previous studies. 
Taking the 2 $\sigma$ upper limit for [O {\footnotesize III}] 
($\approx2.3\times 10^{-18}$ erg s$^{-1}$ cm$^{-2}$), the [O {\footnotesize III}]/[O {\footnotesize II}] ratio of ADF22-QG1 
is $\lesssim 0.4$ corresponding to an H {\footnotesize II} region or LINER \citep{2006MNRAS.372..961K}.
Since the [O {\footnotesize III}] flux upper limit corresponds to 
$\rm\lesssim 2.4\times10^{42}~erg~s^{-1}$, adopting the empirical [O {\footnotesize III}] 
to X-ray luminosity relation for type 2 AGN in \citet{2009A&A...504...73L}, 
it is no wonder that ADF22-QG1 is not detected with the {\it Chandra} observation of the SSA22 protocluster 
with a sensitivity limit $L_{\rm 2-10 keV}\approx5.7\times10^{42}\rm~erg~s^{-1}$ in \citet{2009MNRAS.400..299L}. 
From these above, the [O {\footnotesize II}] emission of ADF22-QG1 may originate in a partly remaining star formation or weak AGN. 
Empirically, the excess [O {\footnotesize II}] of massive quiescent galaxies at high redshift is likely due 
to AGNs or LINERs rather than star formations (e.g., \citealt{2006ApJ...648..281Y,2010ApJ...716..970L}). 

\subsection{SED fitting}\label{subsec:sedfitting}

The absorption features together with multi-wavelength photometry 
are strong constraints on SFHs of massive quenched galaxies  
as demonstrated in previous studies 
(e.g., \citealt{2018A&A...618A..85S,2019ApJ...874...17B,2020ApJ...889...93V,2020ApJ...903...47F,2021ApJ...907L...8A}).
Here we fit the spectrum and photometry of ADF22-QG1 simultaneously 
with the SED models with parametric SFHs by using {\sf FAST++}\footnote{https://github.com/cschreib/fastpp},  
and nonparametric SFHs by using  {\sf Prospector}\footnote{https://github.com/bd-j/prospector} 
\citep{2017ApJ...837..170L, 2017zndo...1116491J}. 

\subsubsection{Data}

First, we prepare the input photometric and spectroscopic data for SED modelings. 
We use the $u^{\star}BVRi'z'JHK_s, 3.6, 4.5, 5.8~\&~8.0~\mu$m-band photometry
measured in \citet{2013ApJ...778..170K}. 
Briefly, we convolve the $u^{\star}$ to $K_s$-band images to match the PSF to a FWHM
of $\approx1.0$ arcsec, and measure fluxes with a 2.0 arcsec diameter aperture.
To match with them, the IRAC $3.6-8.0~\mu$m photometry is applied aperture correction
computed by using the $K_s$-band image (see detail in \citealt{2013ApJ...778..170K}).
We also use the flux measured on the archival $F814W$-band image taken 
with Advanced Camera and Spectrograph on Hubble Space Telescope (PID9760), with the same manner, 
and IR luminosity (at $8-1000~\mu$m; $L_{IR}$) limit 
based on the 1.2 mm (256.98 GHz) image taken with ALMA in Cycle-2 \citep{2018PASJ...70...65U}
and Cycle-5 (PID. 2017.1.01332.S, PI. H. Umehata, H. Umehata et al in prep). 
The $L_{IR}$ is calculated by assuming the average 1.2 mm flux
to $L_{IR}$ relation for the SED library in \citet{2017ApJ...840...78D}. 
ADF22-QG1 is not detected at 1.2 mm 
and the 3$\sigma$ limiting flux is 75 $\mu$Jy corresponding to $L_{IR}\sim0.9-2.0\times10^{11}~L_{\odot}$ taking the 95\% confidence interval.
The input observational data for {\sf Prospector} is the same as that for {\sf FAST++} 
but the former uses $L_{IR}$ while the latter uses the 1.2 mm flux.  

We correct the aperture photometry and one-dimensional spectrum
by multiplying them with 1.21 which is the ratio of the total (Kron) flux
measured on the original image and aperture flux measured on the PSF matched image in $K_s$-band. 
We also correct the Galactic extinction $E(B-V)=0.053$
found from the dust extinction finding tool at NASA/IPAC INFRARED SCIENCE ARCHIVE\footnote{https://irsa.ipac.caltech.edu/applications/DUST/} based on \citet{ 2011ApJ...737..103S}.
The one-dimensional spectrum is re-binned with six pixels ($\approx13$~\AA) 
by taking the inverse variance weighting average.
The [O {\footnotesize II}] flux estimated with {\sf SLINEFIT} is subtracted from the $H$-band photometry. 
The [O {\footnotesize II}] and strong OH airglow on the spectrum are excluded from the SED fitting.
We use the spectrum at $1.51-1.80$ and $1.97-2.19~\mu$m avoiding 
the spectrum with low transmission due to the instrument and/or sky. 
Both transmissions are still good but we avoid using the spectrum at $>2.19~\mu$m
because the non-uniformity of the sky transmission significantly remains on the reduced spectrum. 

\subsubsection{FAST++} \label{subsec:fastpp}
 
First, we fit the spectrum and photometry simultaneously 
with SED models by using {\sf FAST++} following \citet{2018A&A...618A..85S}.
We fit the SED with \citet{2003MNRAS.344.1000B} stellar population synthesis models 
adopting the \citet{2003PASP..115..763C} IMF,  
\citet{2001PASP..113.1449C} dust attenuation law 
and the solar metallicity ($Z=0.02$).
We fit the models with extinction value $A_V=0-6$ mag with steps of 0.1 mag and $\rm \log (age~yr^{-1})=7-9.3$ with steps of 0.05 dex. 
We test SFH of exponentially declining (SFR $\propto \exp (-t/{\tau})$), 
delayed exponentially declining ($\propto t \times \exp(-t/\tau)$), where $t$ is time from the formation, 
and composite SFH in \citet{2018A&A...618A..85S}.
This SFH is described as a combination of,
\begin{equation}
 \rm SFR_{\rm base}(t) \propto \left\{
  \begin{array}{l} 
      e^{(t_{\rm burst}-t) /\tau_{\rm rise}}\rm~for~t > t_{\rm burst}\\
       e^{(t_{\rm burst}-t) /\tau_{\rm decl}}\rm~for~t \leq t_{\rm burst}
    \end{array}
    \right.
\end{equation}
and
\begin{equation}
 \rm SFR(t) =  SFR_{\rm base}(t) \times \left\{
  \begin{array}{l} 
      1\rm~for~t > t_{\rm free}\\
      R_{\rm SFR}\rm~for~t \leq t_{\rm free}
    \end{array}
    \right.
 \end{equation}
\noindent where $t$ is the lookback time. 
We fit in ranges $t_{\rm burst} =$ [10 Myr, $t_{\rm obs}]$ with (logarithmic) steps of 0.05 dex, 
$\rm \tau_{rise},\tau_{decl} =$ [10 Myr, 3 Gyr] with steps of 0.1 dex, 
$t_{\rm free}=$ [10 Myr, 300 Myr] with steps of 0.5 dex, 
and $R_{\rm SFR}=[10^{-2}, 10^5$] with steps of 0.2 dex. 
The velocity dispersion applied for the templates in each {\sf FAST++} run is fixed.
Then we run {\sf FAST++} with a velocity dispersion 
between 100 and 800 km s$^{-1}$ with steps of 20 km s$^{-1}$.

\begin{deluxetable}{lc}
\tablenum{1}
\tablecaption{Physical properties}
\tablehead{ \colhead{Properties} & \colhead{ADF22-QG1}}
\startdata
$z_{\rm spec}$ &  $3.0922^{+0.0008}_{-0.0004}$ \\
$r_{\rm eff}$/kpc$^a$ &  $1.01\pm0.04$ \\
$\log(\rm M_{\star, \rm FAST++}/M_{\odot})$ & $11.26^{+0.03}_{-0.04}$ \\
$\log(\rm M_{\star, continuity}/M_{\odot})$ & $11.54^{+0.03}_{-0.00}$ \\
$\rm SFR_{\rm FAST++}/M_{\odot}~yr^{-1}$ & $<0.3$ \\
$\rm SFR_{\rm continuity}/M_{\odot}~yr^{-1}$ & $0.01^{+0.03}_{-0.01}$ \\
$\rm SFR_{\rm [O~II]}/M_{\odot}~yr^{-1}$ & $2-14$ \\
$\rm SFR_{\rm IR}/M_{\odot}~yr^{-1}$ & $<9-21$\\
$\rm sSFR_{\rm FAST++}/yr^{-1}$ & $<1.8\times10^{-12}$ \\
$\rm sSFR_{\rm continuity}/yr^{-1}$ & $0.7^{+10.3}_{-0.7}\times10^{-14}$ \\
$\rm sSFR_{\rm [O~II]}/yr^{-1}$$^b$ & $1-8\times10^{-11}$ \\
$\rm sSFR_{\rm IR}/yr^{-1}$$^b$ & $<5-12\times10^{-11}$\\
$A_{V, \rm FAST++}$/mag & $0.5^{+0.1}_{-0.1}$ \\
$\hat{\tau}_{\rm2, continuity}$ & $0.13^{+0.02}_{-0.02}$ \\
$\log(t_{50, \rm FAST++}$/yr) & $8.79^{+0.05}_{-0.05}$ \\
$\log(t_{\rm 50, continuity}$/yr) & $9.08^{+0.08}_{-0.02}$ \\
$\log(\rm \langle SFR \rangle_{\rm main, FAST++}/M_{\odot}~yr^{-1})$ & $4.06^{+0.45}_{-1.23}$ \\
$\log(\rm \langle SFR \rangle_{\rm main, continuity}/M_{\odot}~yr^{-1})$ & $2.31^{+0.06}_{-0.04}$ \\
$\log(t_{q, \rm FAST++}$/yr) & $8.78^{+0.04}_{-0.20}$ \\
$\log(t_{q, \rm continuity}$/yr) & $8.80-8.91$ \\
$\log(\tau_{\rm decl, FAST++}\rm/yr$) & $7.0^{+2.5}_{-0.0}$ \\
\enddata
\footnote{\citet{2017MNRAS.469.2235K}}
\footnote{Adopting M$_{\star, \rm FAST++}$ as an upper limit value.}
\end{deluxetable}

\begin{figure*}
\gridline{\fig{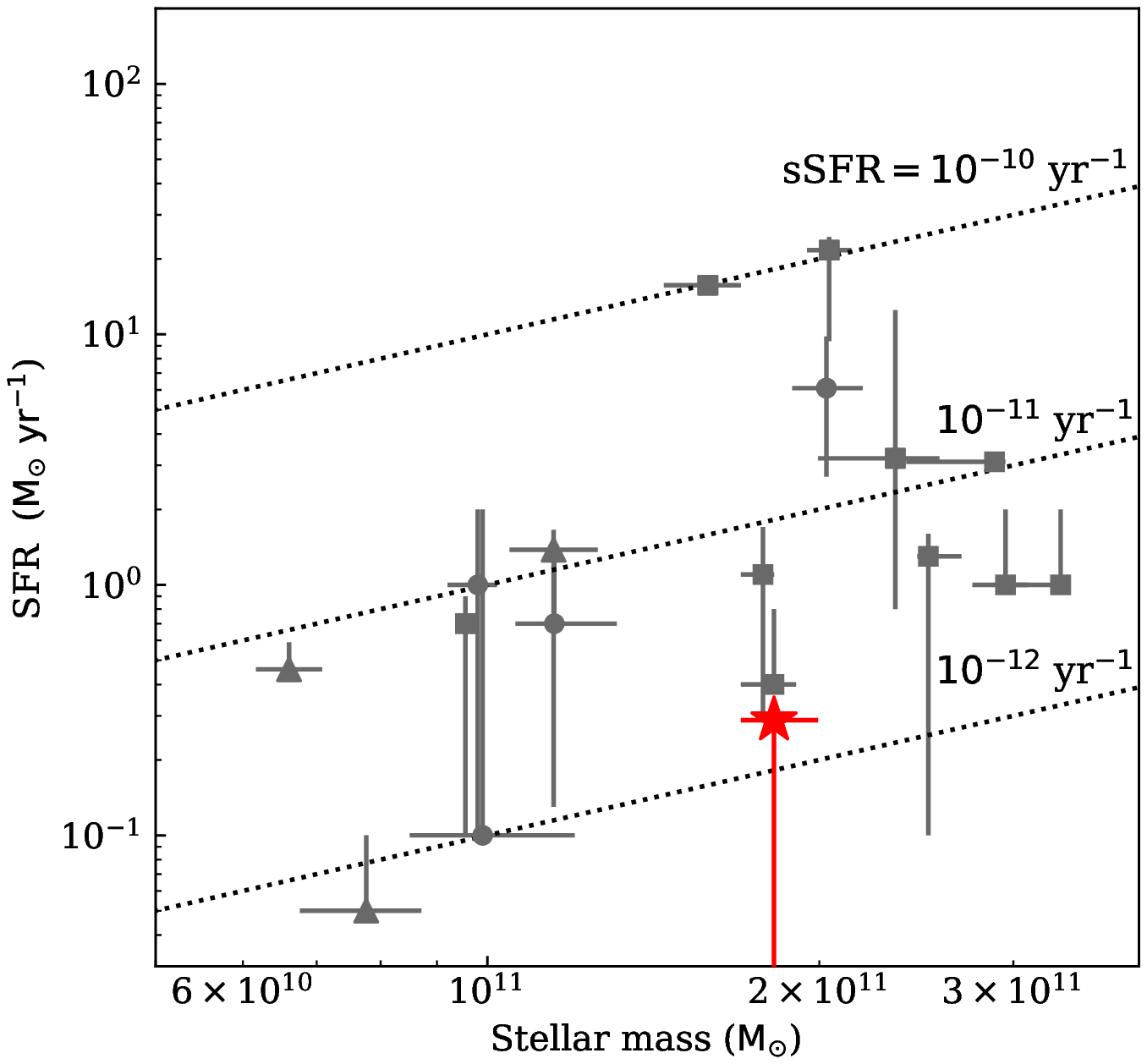}{0.33\textwidth}{}\fig{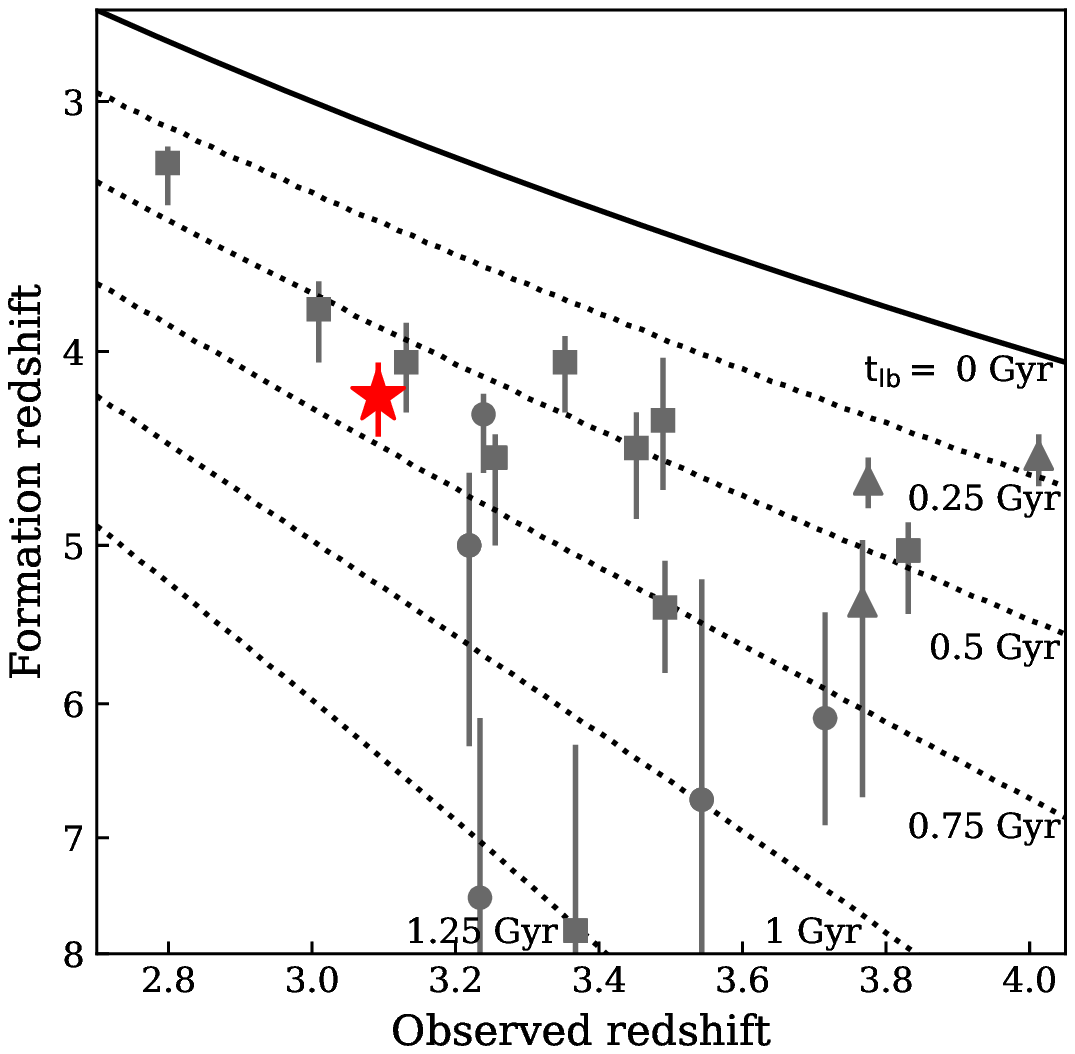}{0.33\textwidth}{}\fig{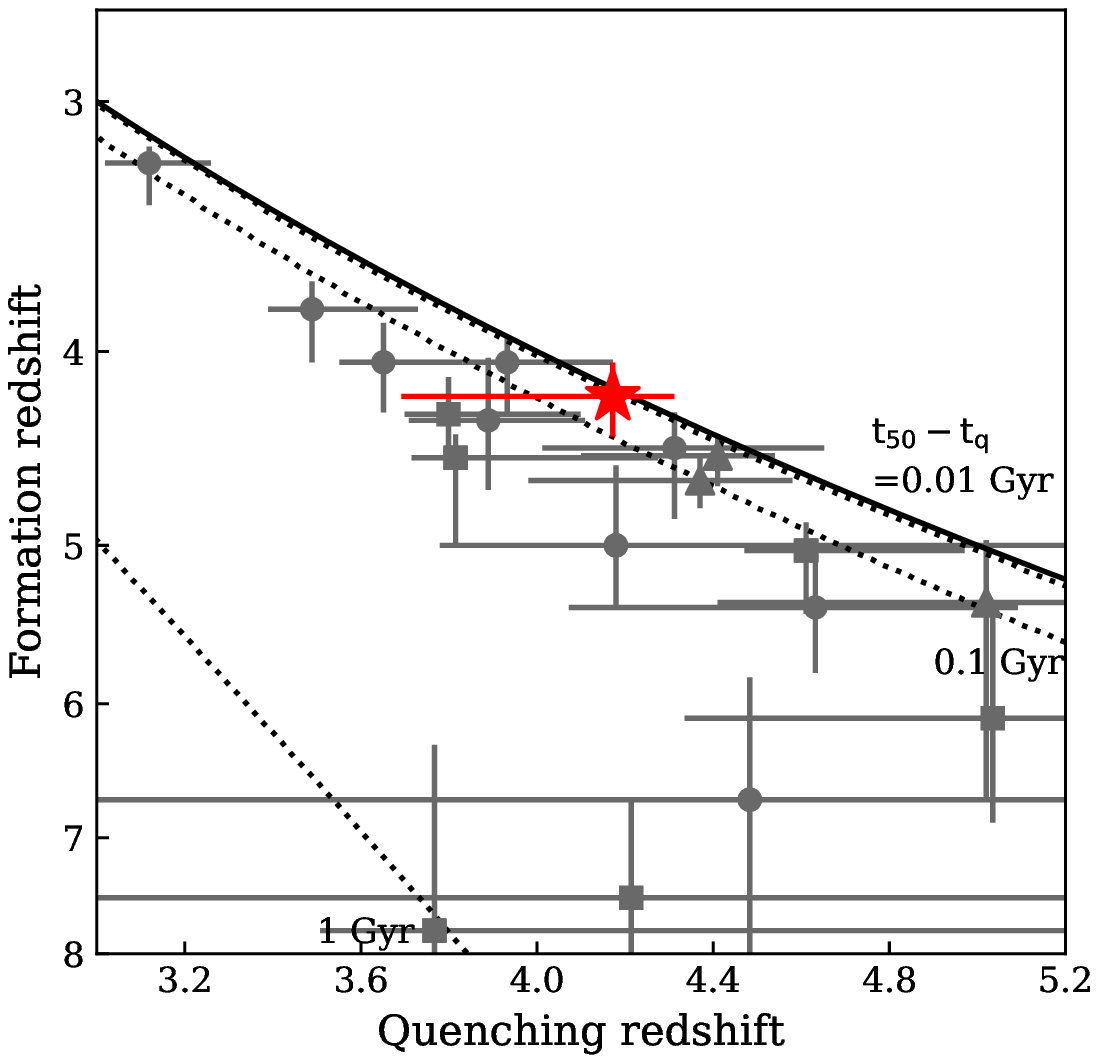}{0.33\textwidth}{}}
\caption{The comparison of the SED quantities. 
The red star shows ADF22-QG1.
The gray circles, squares, and triangles show quenched galaxies
in \citet{2018A&A...618A..85S}, \citet{2020ApJ...903...47F}, and \citet{2020ApJ...889...93V}.
{\it Left}: The stellar mass vs. SFR.  
The dotted curves show the sSFR of $10^{-10}$ yr$^{-1}$, $10^{-11}$ yr$^{-1}$ and $10^{-12}$ yr$^{-1}$.
{\it Center}: The observed redshifts vs. formation redshifts. 
The dotted curves show the lookback times from the observed redshifts with 0.25 Gyr steps. 
{\it Right} The quenching redshifts vs. formation redshifts.
The dotted curves show the time durations between formation and quenching redshifts of $0.01, 0.1$ and 1 Gyr. 
\label{fig:zfzqzobs}}
\end{figure*}

\begin{figure}
\gridline{\fig{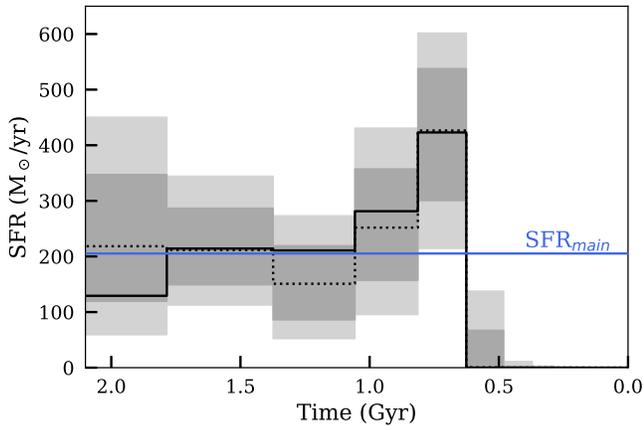}{0.5\textwidth}{}}
\caption{The SFH evaluated adopting the continuity SFH prior with {\sf Prospector}.
The Black solid histogram shows the MAP value. 
The gray shaded region and dotted line show the 16, 50, and 84th percentiles of the posterior.  
The light gray shaded region show the 5 and 95th percentiles of the posterior.  
\label{fig:sfh}}
\end{figure}

\subsubsection{Prospector}

A parametric SED modeling can fail to reproduce a true SFH 
if a galaxy has a more complex SFH, e.g., starburst, sudden quenching and rejuvenation. 
Then we also fit the SED with nonparametric models which can handle 
with complex SFH following \citet{2017ApJ...837..170L, 2019ApJ...876....3L}. 
The fitting is performed by using {\sf Prospector} \citep{2017ApJ...837..170L, 2017zndo...1116491J} 
which uses the Flexible Stellar Population Synthesis ({\sf FSPS}) code \citep{2009ApJ...699..486C}. 
As in the {\sf FAST++} SED modeling, the \citet{2003PASP..115..763C} IMF 
and the \citet{2001PASP..113.1449C} dust attenuation law  are adopted.
The two-component \citet{2000ApJ...539..718C} dust attenuation model is adopted  in {\sf Prospector}
while the uniform screen model is adopted in {\sf FAST++}. 
It consists from birth-clouds and diffuse dust screen components. 
The dust attenuation is parameterized with the optical depths at 5500 \AA~of these two components,  
$\hat{\tau}_1$ (birth-clouds) and $\hat{\tau}_2$ (diffuse dust). 
In case of the \citet{2001PASP..113.1449C} dust attenuation law, 
$\hat{\tau}_1$ is set to zero and the dust attenuation is controlled only with $\hat{\tau}_2$. 
We adopt the result measured adopting the solar metallicity ($Z=0.02$).

We use a continuity SFH prior which fits directly for $\Delta\log$(SFR) between adjacent time bins. 
Here the adopted time bins are  $0<\log(\rm age/yr)<7.5$, $7.5<\log(\rm age/yr)<8.0$, 
and  $\Delta\log(\rm age/yr)=0.11$ binning at the range $8.0<\log(\rm age/yr)<9.3$ (14 time bins in total). 
The free parameters are $\log(M_{\star}/M_{\odot})$ (between 7 to 12), $\Delta\log$(SFR) 
between adjacent time bins (scale =0.3, df =2),  and $\hat{\tau}_2$ (between 0.0 to 2.0). 
The sampling is performed with the nested sampler {\sf dynesty} \citep{2020MNRAS.493.3132S} 
and the maximum a posteriori (MAP) values are presented as the best-fit quantities. 

Though we adopt the above model, 
we also test the continuity SFH model with free metallicity, and prospector-$\alpha$ \citep{2017ApJ...837..170L} model. 
The latter model adopts a Dirichlet SFH prior, the prescription of dust attenuation law 
from \citet{2009A&A...507.1793N} which allows a flexible dust attenuation law slope 
and $\hat{\tau}_1$ to $\hat{\tau}_2$ ratio, 
and adds an AGN contribution. 
We summarize the results of these different modeling in Appendix B. 
Briefly, the results do not depend on the dust attenuation law and the contribution of AGN is negligible. 
The model with near solar metallicity is favored for the continuity-SFH model 
but low metallicity is favored for the {\sf Prospector}-$\alpha$ (Dirichlet) model, 
though our data has no significant spectral indices to determine the metallicity definitely. 
A model with lower metallicity tends to have a SFH in which the star formation quenches earlier.
Thus we here adopt the continuity-SFH model with the solar metallicity ($Z=0.02$) as the conservative model.

\section{Result}\label{sec:Result}

\subsection{SED: FAST++}\label{subsec:SED}

The red curves in Fig. \ref{fig:spectra} and \ref{fig:phot} show the best-fit SED, 
and Table 1 lists the SED parameters and SFH quantities found with {\sf FAST++}.
We note that there is no significant difference in the SED quantities and SFHs derived with the three SFH models. 
Here we adopt the best-fit model found with the composite SFH modeling
to compare the results directly with literature \citep{2018A&A...618A..85S,2020ApJ...889...93V,2020ApJ...903...47F}.
The SFH and several SFH quantities, $\langle SFR \rangle_{\rm main}$, $t_{50}$ and $t_{q}$, are computed with {\sf FAST++}. 
The $\langle SFR \rangle_{\rm main}$ is the mean SFR during the shortest time interval 
over which 68 \%~of the star formation took place. 
The formation time $t_{50}$ is the time at which 50\% of the total stellar mass has been formed, excluding mass loss and recycling, 
and the quenching time $t_q$ is the elapsed time since the SFR dropped below 10\% of the $\langle SFR \rangle_{\rm main}$. 
Both $t_{50}$ and $t_{q}$ are given as a lookback time from the observed redshift. 
The uncertainties of the values from the SED modeling in Table 1 
are the 90\% confidence interval values. 

The best-fit model and its reduced $\chi^2$ value change marginally 
by setting different velocity dispersion. 
We also tried the penalized pixel-fitting algorithm (pPXF; \citealt{2017MNRAS.466..798C})
but could not also obtain a robust velocity dispersion estimate. 
It is because all the available Balmer absorption features are partly covered by OH airglow. 
Here we show the best-fit parameters for velocity dispersion 
$=320$ km s$^{-1}$ where the $\chi^2$ value minimizes ($\chi^2/\nu=2.7$). 
We also test the models with metallicities 0.2, 0.4 and 2.5 $Z_{\odot}$ 
but it also do not change the results significantly. 
A corner plot for covariance among the age, $A_{\rm V}$, Z, SFR, $M_{\star}$, $t_{50}$ and $t_{q}$ 
is presented in Fig. \ref{fig:cornerplot}, which shows 
that the stellar population of ADF22-QG1 is well constrained with small degeneracy. 

ADF22-QG1 is well fitted with a SED model suppressed star formation 
of the specific SFR $\rm (sSFR_{\rm FAST++}) <1.8\times10^{-12}$ yr$^{-1}$. 
The $L_{\rm IR}$ limit for ADF22-QG1 corresponds to 
SFR $\rm<9-21~M_{\odot}$ yr$^{-1}$ using the $L_{\rm IR}$ to SFR 
conversion factor in \citet{2012ARA&A..50..531K}. 
Note that the $L_{\rm IR}$ and thus $\rm SFR_{\rm IR}$ based on a single-band flux 
depend greatly on the adopted SED models \citep{2019MNRAS.486..560S,2021arXiv210508894M}. 
Scaling the 1.2 mm flux limit of our target to the model prediction in \citet{2019MNRAS.486..560S}, 
it can have a SFR$_{\rm IR}$ upper limit $\rm\lesssim100~M_{\odot}~yr^{-1}$. 
Even though we adopt the $\rm SFR_{\rm [O~II]}$ or $\rm SFR_{\rm IR}$ at upper limit, 
ADF22-QG1 is classified as a quenched galaxy. 
We put the SFH evaluated with  {\sf FAST++} in Fig. \ref{fig:sfhfastpp} 
where the whole stellar mass is formed instantaneously.
The formation time $t_{50}$ of ADF22-QG1 is younger ($0.62^{+0.09}_{-0.05}$ Gyr) 
than that estimated from the fitting only with the photometry ($\gtrsim 1$ Gyr).  
This is expected from the spectral characteristic of ADF22-QG1 
that 4000 \AA~break is strong but Balmer absorption features are still significant.
Such a SFH cannot be measured robustly without deep NIR spectroscopic observations like this study. 

Figure \ref{fig:zfzqzobs} compares the SED quantities of ADF22-QG1, 
and quenched galaxies at $2.8<z_{\rm spec}<4.0$ 
in literature \citep{2018A&A...618A..85S,2020ApJ...889...93V,2020ApJ...903...47F}. 
Star formation of ADF22-QG1 is suppressed well among them. 
ADF22-QG1 has formed later but more rapidly than other QGs. 
At this point, we find no significant environmental dependance in the SED quantities. 
Note that the figure looks lacking slowly quenched ($t_{50}-t_{q}>0.1$ Gyr) galaxies 
however this is likely the sample bias that quiescent galaxies at $z>3$  quenched earlier 
than the current sample are hardly surveyed completely 
at the current survey depth $K_s \lesssim 24$ mag  \citep{2020ApJ...903...47F}.

\subsection{SED: nonparametric}\label{subsec:SED-nonparam}

The best-fit SED from nonparametric modeling of SFH is shown 
with blue curves in Fig. \ref{fig:spectra} and \ref{fig:phot}.
It also fits the observed SED well. 
The SED quantities from nonparametric modeling are denoted with continuity in Table 1. 
Fig. \ref{fig:cornerplot_continuity} shows a corner plot 
for covariance among the fitting parameters in the nonparametric modeling. 

Both the nonparametric and parametric modeling fit the observed SED 
with a model of massive galaxy with suppressed star formation ($\rm sSFR_{\rm continuity} = 0.7^{+10.3}_{-0.7}\times10^{-14}~yr^{-1}$).
However, they find different SFHs and stellar masses. 
Fig. \ref{fig:sfh} shows the SFH evaluated with nonparametric modeling. 
It is flatter than that from {\sf FAST++} and does not take an extreme value.
The SFR drops sharply after the time bin $8.80<\log(\rm age/yr)<8.91$ 
or the star formation is quenched during this time bin. 
It agrees very well with the quenching time $\rm\log(t_{q, \rm FAST++}/yr) = 8.78^{+0.04}_{-0.20}$ from {\sf FAST++}.  
Because of the larger contribution from old stars, 
the stellar mass from nonparametric modeling is twice larger than that from {\sf FAST++}. 

\subsection{Newly detected [O {\footnotesize III}] emitters}

\begin{figure}
\gridline{\fig{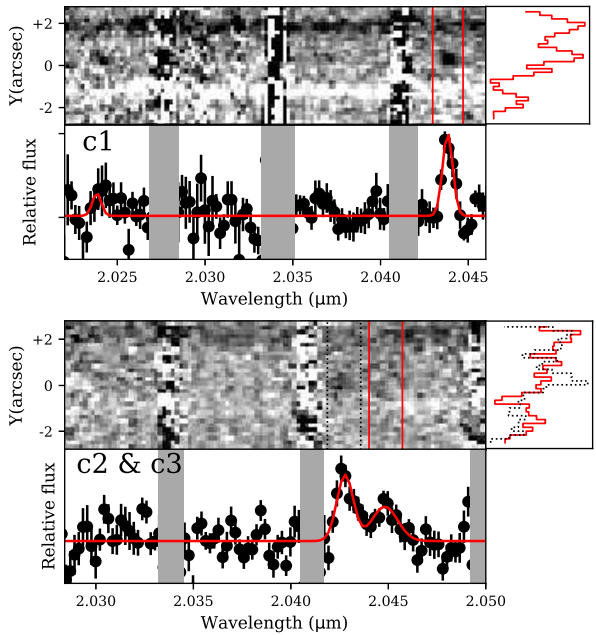}{0.45\textwidth}{}}
\caption{{\it Top:} The spectrum at c1. The {\it left} panel shows the spectrum image. 
The {\it bottom} panel shows the one-dimensional spectrum. 
Black points show the observed spectrum. 
The gray shaded regions show OH airglow masked when we fit the spectrum.
The red curve shows the Gaussian fit of the spectrum. 
The {\it top right} shows the spatial profile of the spectrum measured 
between the red solid lines in the {\it bottom} panel. 
{\it Bottom:} Similar to the {\it top} panel but for c2 (redder) and c3 (bluer). 
The spatial profiles of the spectra are shown at two wavelength ranges corresponding to c2 and c3 (red solid and black dashed lines). 
\label{fig:oiii_spectra}}
\end{figure}

Adjacent to ADF22-QG1, emission lines are detected 
at three positions between $2.042$ and $2.045$ $\mu$m (Fig. \ref{fig:oiii_spectra}, here after c1, c2 and c3). 
We extract their one-dimensional spectra by combining 
the spectra for $\approx1.3$ and 2.2 arcsec in the spatial direction 
for c1, and c2 \& c3, respectively, 
and smoothing them with three pixels in the wavelength direction. 
We fit the one-dimensional spectra with Gaussian profiles 
by a standard $\chi^2$ minimization procedure.
We derive them under the assumption
that they are  [O {\footnotesize III}]$\lambda$5007.

We estimate their sky positions based on 
the average spatial profiles in a range $\approx17$ \AA~(8 pixels) at emission lines 
(Fig. \ref{fig:oiii_spectra} {\it right tops}). 
In the case of c1 and c3, they show clear peaks at 1.6 and 2.3 arcsec from ADF22-QG1. 
The peak of c2 is not so clear but likely fallen between c3 and ADF22-QG1. 
Their $K_s$-band magnitudes measured with 0.80 arcsec diameter apertures are 
$<26.34~(<2\sigma$), $26.02~(2.7\sigma)$ and $25.72~(3.5\sigma)$, respectively. 
Table 2 summarizes their observed properties. 
Interestingly, the spatial locations and redshifts of c2 and c3 are just between ADF22-QG1 and the nearest massive galaxy, K15d 
at $z_{\rm spec}=3.0774\pm0.0003$ \citep{2016MNRAS.455.3333K}. 

\begin{deluxetable*}{lccccc}
\tablenum{2}
\tablecaption{Newly detected [O {\footnotesize III}] sources}
\tablehead{ \colhead{ID} & \colhead{R.A.} & \colhead{Dec}& \colhead{$z_{\rm spec}$}& \colhead{$F_{\rm[OIII]}$}& \colhead{$K_s$}\\
\colhead{} & \colhead{[J2000]}& \colhead{[J2000]}& \colhead{} & \colhead{($10^{-18}$ erg cm$^{-2}$)}& \colhead{(mag)} }
\startdata
c1 & 22:17:37.24 & +0:18:14.2 & $3.0811\pm0.0001$ & $0.9\pm 0.2$  &$<26.34^a$ \\
c2 & 22:17:37.17 & +0:18:16.7 & $3.0833\pm0.0003$ & $0.9\pm0.2$ & $26.02$ \\
c3 & 22:17:37.13 & +0:18:17.1 & $3.0791\pm0.0001$ &  $1.1\pm0.2$ & $25.72$ \\
\enddata
\footnote{2$\sigma$ limiting flux}
\end{deluxetable*}

\section{Discussion}\label{sec:discussion} 

\subsection{SFH}

Both the parametric and nonparametric modeling of the SFH
find that ADF22-QG1 is rapidly quenched at $\sim0.6$ Gyr ago.
To reproduce such a sudden quenching, 
strong feedbacks are needed (e.g., \citealt{2019ApJ...874...17B, 2019MNRAS.490.2139R, 2020ApJ...890L...1F}). 
If the  SFH of ADF22-QG1 is alike with that estimated with {\sf FAST++}, 
feedback from young massive stars is applicable for the quenching \citep{2005ApJ...618..569M}.
The quenching via AGN feedback cannot be ignored especially for massive galaxies.
ADF22-QG1 is not detected with {\it Chandra} but its [O {\footnotesize II}] 
can originate in a dying AGN. 
We will further discuss the roles of AGNs for the evolution of protocluster galaxies in an upcoming paper. 
It is beyond the scope of this paper but also a problem to be solved in the future that 
how ADF22-QG1 has maintained its quiescence for several 100 Myr
although it is in a gas rich environment and surrounded 
by star-bursting neighbors \citep{2019Sci...366...97U} which can induce further starburst. 

Which SFH is more realistic?
The $\langle SFR \rangle_{\rm main, FAST++}$ of ADF22-QG1 
is not so realistic because it is even higher than those of the brightest SMGs
with SFR $\lesssim 5000$ M$_{\odot}$ yr$^{-1}$ 
\citep{2013Natur.496..329R, 2016ApJ...827...34O,2020ApJ...905...86S}. 
Theoretically, if the SFR surface density of a SFG reaches the Eddington limit 
where momentum-driven wind induced by radiation pressure on dust heated by young massive stars, 
its star formation can be suppressed  \citep{2005ApJ...618..569M}.
The limiting SFR surface density for SMGs is $\rm\sim1000~M_{\odot}$ yr$^{-1}$ kpc$^{-2}$
based on \citet{2011ApJ...727...97A}, 
and SMGs do not exceed this limit in general \citep{2015ApJ...807..128S} while the brightest SMGs are near this limit. 
The SFR surface density of ADF22-QG1 at $\langle SFR \rangle_{\rm main, FAST++}$ 
computed by dividing it with an area with a radius of the twice $r_{eff}$ at the observed redshift 
is $0.86^{+1.67}_{-0.78}\times10^3$ M$_{\odot}$ yr$^{-1}$ kpc$^{-2}$ 
which by far exceeds the Eddington limit SFR surface density. 

Thus the SFH from nonparametric modeling is more realistic.
The SFR of ADF22-QG1 before quenching is consistent 
with or lower than those of the observed SMGs in protoclusters 
at $z\sim4$ (e.g., \citealt{2018ApJ...856...72O,2018ApJ...861...43P, 2018Natur.556..469M,2020ApJ...898..133L})
and the SSA22 protocluster itself \citep{2018PASJ...70...65U}. 
Our argument can also be true for the other massive quiescent galaxies at $z>3$ 
in literature in which parametric SED modelings are widely used.
In these studies, the SFHs are often described as a vigorous starburst 
similar to ADF22-QG1 computed with {\sf FAST++}
(e.g., \citealt{2018A&A...618A..85S,2020ApJ...889...93V,2020ApJ...903...47F}). 
Then the most active starburst galaxies like SMGs have SFRs in concordance with the SFH for massive quiescent galaxies. 
However, to explain the observed properties, e.g., number densities, 
less-bursty SFGs on the star formation main sequence are also needed to be the major progenitors of massive quiescent galaxies 
\citep{2013ApJ...765..104B,2017ApJ...840...47B,2017A&A...602A..11P,2019ApJ...886...88G,2020ApJ...889...93V}. 
In addition, observed stellar masses of the brightest SMGs at $z>3$ 
are often larger than that of massive quiescent galaxies at $3<z<4$ \citep{2020ApJ...889...93V}. 
Adopting the SFH from nonparametric modeling, 
massive SMGs and main sequence SFGs are allowed as a progenitor of ADF22-QG1.
Although we need a further robust SFH measurement in the future, 
application of SFHs more complex than typical parametric SFHs
can correct the relation between massive quiescent galaxies and high-z SFGs. 

\begin{figure}
\gridline{\fig{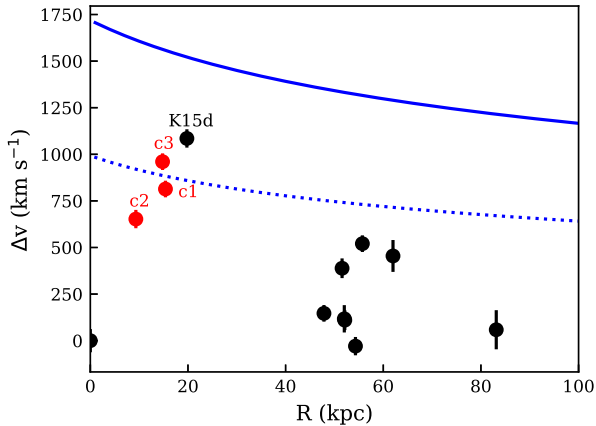}{0.45\textwidth}{}}
\caption{The line-of-sight velocities and spatial distances of the galaxies relative to the ADF22-QG1. 
The black points are based on the redshifts of galaxies listed in  \citet{2016MNRAS.455.3333K} and \citet{2019Sci...366...97U} 
while the red points show the [O {\footnotesize III}] emitters confirmed in this study. 
The blue solid curve show the escape velocity for a NFW halo with a halo mass $1.1\times10^{13}$ M$_{\odot}$  
(virial radius $=170$ kpc and concentration $c=5$). 
The blue dashed curve is that scaled to projected one (the velocity and distance are divided with $\sqrt3$ and $\sqrt{3/2}$, respectively). 
\label{fig:velocity_dist}}
\end{figure}

\subsection{Size and mass evolution}

Here we confirm a compact massive quiescent galaxy as a secure progenitor 
of a giant elliptical or brightest cluster galaxy (BCG) in a cluster of galaxies today. 
Similarly, they reported that massive quiescent galaxies in proclusters/clusters at $z\lesssim2$
are more compact than nearby giant ellipticals in \citet{2008ApJ...680..224Z, 2013ApJ...772..118S,2015ApJ...804..117M}. 
It needs a strong size evolution to evolve into a typical giant elliptical or BCG. 
Cosmological numerical simulations predict that a BCG is hierarchically 
formed via multiple mergers of galaxies (e.g., \citealt{2007MNRAS.375....2D}). 
Supporting such a scenario, ADF22-QG1 is not isolated 
but in a dense group of massive galaxies and SMGs. 
Fig. \ref{fig:velocity_dist} shows the velocity distribution of galaxies within 100 physical kpc from ADF22-QG1. 
We use the redshifts derived from [O {\footnotesize III}] and/or CO(3-2) lines 
in \citet{2016MNRAS.455.3333K} and \citet{2019Sci...366...97U} (black), 
and also newly confirmed [O {\footnotesize III}] emitters (red) in this study.
The redshift of ADF22-QG1 is certainly close to the group members though 
it is offset from the median redshift $z_{\rm med}=3.087$ of the group. 
Their velocity dispersion is $\rm\sigma_v=351\pm52$ km s$^{-1}$ evaluated by a bootstrap resampling. 
According to the scaling relation based on $N$-body simulations in \citet{2008ApJ...672..122E}, 

$$\sigma_{\rm DM}(M,z) =  \sigma_{\rm DM,15} \left[ \frac{h(z)M_{200}}{10^{15} \rm M_{\odot}} \right]^\alpha$$
where $h(z)=H(z)$/100 km s$^{-1}$ and $\sigma_{\rm DM,15}$ 
is the normalization for a halo mass $10^{15}~h^{-1}$ M$_{\odot}$.
They found $\sigma_{\rm DM,15}=1082.9\pm4.0$ km s$^{-1}$ 
and $\alpha=0.3361\pm0.0026$ for $\Lambda$CDM cosmology. 
Adopting the above $\sigma_v$, the halo mass of AzTEC14 group is $1.1\pm0.4\times10^{13}$ M$_{\odot}$
consistent with our previous measurement \citep{2016MNRAS.455.3333K}. 
The velocity offsets of all the group members are below the escape velocity 
for a halo characterized with the above halo mass and NFW \citep{1997ApJ...490..493N} profile. 

The velocity distribution and halo mass of AzTEC14 group 
are similar to those of SPT2349–56 which is known as an extreme overdensity of SMGs at $z=4.3$
\citep{2018Natur.556..469M,2020MNRAS.495.3124H}.
\citet{2021MNRAS.502.1797R} shows that the stellar mass 
of the most massive member of $3.2^{+2.3}_{-1.4}\times10^{11}$ M$_{\odot}$
and the total stellar mass of ($12.2\pm2.8)\times10^{11}$ M$_{\odot}$ at least for SPT2349–56.  
Correcting the total stellar mass of the $K_s$-band detected galaxies of AzTEC14 group 
found in \citet{2016MNRAS.455.3333K} with the newly measured stellar mass for ADF22-QG1 
from the nonparametric SFH modeling, 
its total stellar mass is $6.7^{+2.3}_{-0.8}\times10^{11}~M_{\odot}$ at least,
which is comparable or lower than SPT2349–56. 
On the other hand, SPT2349–56 is by far luminous at sub-mm than the SSA22 protocluster \citep{2018Natur.556..469M}, 
i.e., the star formation in AzTEC14 group is less active. 
The SPT2349–56  is a close analog of AzTEC14 group
but a larger stellar mass will be assembled in the former. 

\citet{2020MNRAS.493.4607R} simulated the assembly history of BCGs
by performing isolated non-cosmological hydrodynamical simulations
based on the observed properties of SPT2349–56. 
They predicted that all the group members at $z=4.3$ merge into one massive galaxy by $\sim0.5$ Gyr.
The spectroscopic confirmation of ADF22-QG1 in AzTEC14 group and newly confirmed [O {\footnotesize III}] emitters,
which are possible evidence of interactions between ADF22-QG1 and K15d, 
further support such an early BCG assembly scenario via multiple mergers. 
We discussed the size and stellar mass growths of ADF22-QG1 via mergers 
based on the observed sizes and stellar masses of the AzTEC14 group members in \citet{2017MNRAS.469.2235K}. 
Assuming no further stellar mass and size growths in each group member, 
and all mergers are dry, the size and stellar mass of ADF22-QG1 can be $3-4$ times 
and double from them at the observed redshift via mergers of all the group members, respectively.
If ADF22-QG1 is just a progenitor of a giant elliptical, this scenario is enough. 
However, to grow ADF22-QG1 into a BCG, 
further size and stellar mass growths in each group member 
and/or further mergers of external galaxies are required.
This argument does not change greatly by 
adopting the newly measured stellar mass for ADF22-QG1 in this study. 
Consideration of the newly confirmed SMGs in \citet{2019Sci...366...97U}, 
the size and stellar mass growths especially for them, 
and stellar/AGN feedbacks in AzTEC14 group need further studies in the future. 
Finally we note that the candidate progenitors of the BCG of the SSA22 protocluster 
is not only ADF22-QG1 since there are several such extremely 
dense groups of galaxies \citep{2016MNRAS.455.3333K,2018PASJ...70...65U}.

\section{Conclusion}

We confirm a massive quiescent galaxy at the core of a protocluster at $z=3.09$ in the SSA22 field. 
This is the most distant quiescent galaxy confirmed with Balmer absorption features in a protocluster as ever, 
and a securely selected giant elliptical/BCG progenitor. 
We fit the observed SED with both parametric and nonparametric models of SFHs. 
Both models agree with that our target is a massive galaxy with a suppressed star formation.
The SFH found with the parametric modeling is described as a short starburst 
while that of the nonparametric modeling is a longer-lived SFH.
The SFH found with the nonparametric modeling is more realistic 
given the extremely high SFR surface density at past derived from parametric modeling. 
On the other hand, both models support that the star formation is suddenly quenched after a starburst at $\sim0.6$ Gyr ago. 
To reproduce this, a strong feedback is required. 
This massive quiescent galaxy is confirmed as a member of an extremely dense group of massive galaxies and SMGs 
predicted as a progenitor of a BCG in cosmological numerical simulations. 
According to the simulations, such a system merge into one massive galaxy within $\sim0.5$ Gyr. 
We also newly confirm three plausible [O {\footnotesize III}] emitters around this quiescent galaxy. 
Two of them are possible evidence of the interaction 
between the quiescent galaxy and its nearest massive galaxy. 
They strongly support a hierarchical formation scenario of BCGs. 

It is still unclear what is the major formation and quenching mechanism, 
and how to maintain the quiescence for several 100 Myr of our target in such a dense environment. 
Future studies of older and fainter quiescent galaxies,  
and stellar (gas and star formation) surface density of galaxies 
which may correlate with the SFH and quenching mechanism, 
and the roles of AGNs in protoclusters
with {\it James Webb Space Telescope} ($JWST$) 
and {\it Nancy Grace Roman Space Telescope} ($NGRST$) 
will enable us to discuss the typical formation scenario of cluster galaxies. 

\acknowledgments

We thank the anonymous referee for a number of useful suggestions.
This work is supported by JSPS KAKENHI Grant Numbers 
20K14530 (MK), 17K14252, and 20H01953 (HU).
The spectroscopic data were obtained at the W. M. Keck Observatory, 
which is operated as a scientific partnership among the California Institute of Technology, 
the University of California and the National Aeronautics and Space Administration.
The observations were carried out within the framework of Subaru- Keck/Subaru-Gemini 
time exchange program which is operated by the National Astronomical Observatory of Japan.
The $K_s$-band image was collected with nuMOIRCS at Subaru Telescope, 
which is operated by the National Astronomical Observatory of Japan.
We are honored and grateful for the opportunity of observing the 
Universe from Maunakea, which has the cultural, historical and natural 
significance in Hawaii.
This paper makes use of the following ALMA data: ADS/JAO.ALMA\#2013.1.00162.S, 
ADS/JAO.ALMA\#2016.1.00580.S,
ADS/JAO.ALMA\#2017.1.01332.S. 
ALMA is a partnership of ESO (representing its member states), NSF (USA) and NINS (Japan), together with NRC (Canada), MOST and ASIAA (Taiwan), and KASI (Republic of Korea), in cooperation with the Republic of Chile. The Joint ALMA Observatory is operated by ESO, AUI/NRAO and NAOJ.
The $F814W$-band image is 
based on observations made with the NASA/ESA Hubble Space Telescope, 
obtained from the data archive at the Space Telescope Science Institute. 
STScI is operated by the Association of Universities for Research in Astronomy, 
Inc. under NASA contract NAS 5-26555.

\appendix

\section{UVJ color diagram}
\setcounter{figure}{0}    
\counterwithin{figure}{section}
\begin{figure}
\gridline{\fig{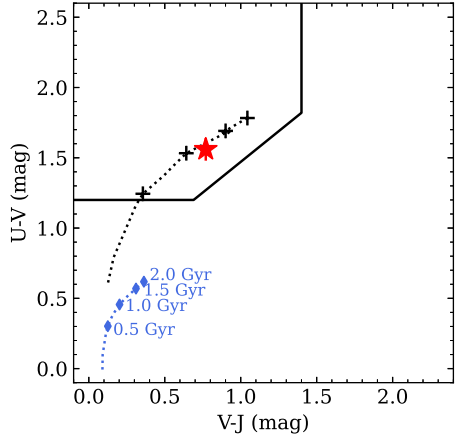}{0.35\textwidth}{}}
\caption{ The rest-frame $UVJ$ color diagram for ADF22-QG1. 
The red star shows the rest-frame $UVJ$ colors of ADF22-QG1 estimated with the SED fitting. 
The points and curves show the color evolution tracks for SED models 
with age between 0.1 to 2  Gyr computed with {\sf GALAXEV} \citep{2003MNRAS.344.1000B}. 
The black crosses with dotted curve show the color evolution track 
for a single burst star formation model with $A_V=0$. 
The blue diamonds and curve show the color evolution track 
for a constant star formation model with $A_V=0$. 
The points are shown at ages 0.5, 1.0, 1.5, and 2.0 Gyr. 
The black solid line show the color criterion for quiescent galaxies at $2.0<z<3.5$, 
$(U-V) > 0.88\times(V -J)+0.59$, $(U-V)>1.2$ and $(V-J)<1.4$ in \citet{2013ApJ...770L..39W}. 
\label{fig:uvj}}
\end{figure}

\section{The corner plots and SFHs for various models}

\begin{figure}
\gridline{\fig{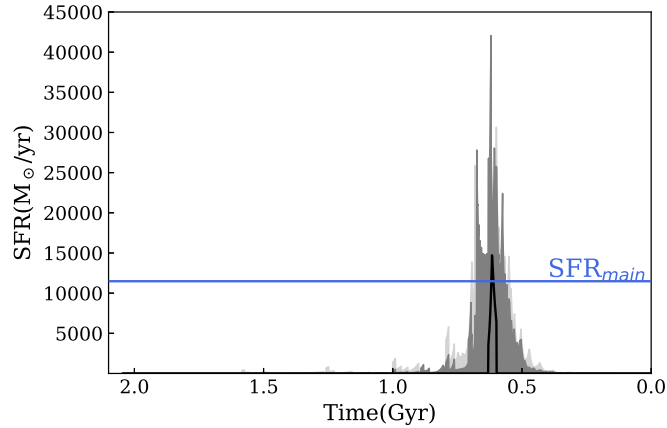}{0.5\textwidth}{}}
\caption{The  SFH evaluated with {\sf FAST++}. 
The black curve show the best-fit SFH.
The gray and light gray shaded region show the 68\% and 90\% confidence interval, respectively. 
The blue vertical line shows the $\langle SFR \rangle_{\rm main}$. 
\label{fig:sfhfastpp}}
\end{figure}

\begin{figure*}
\gridline{\fig{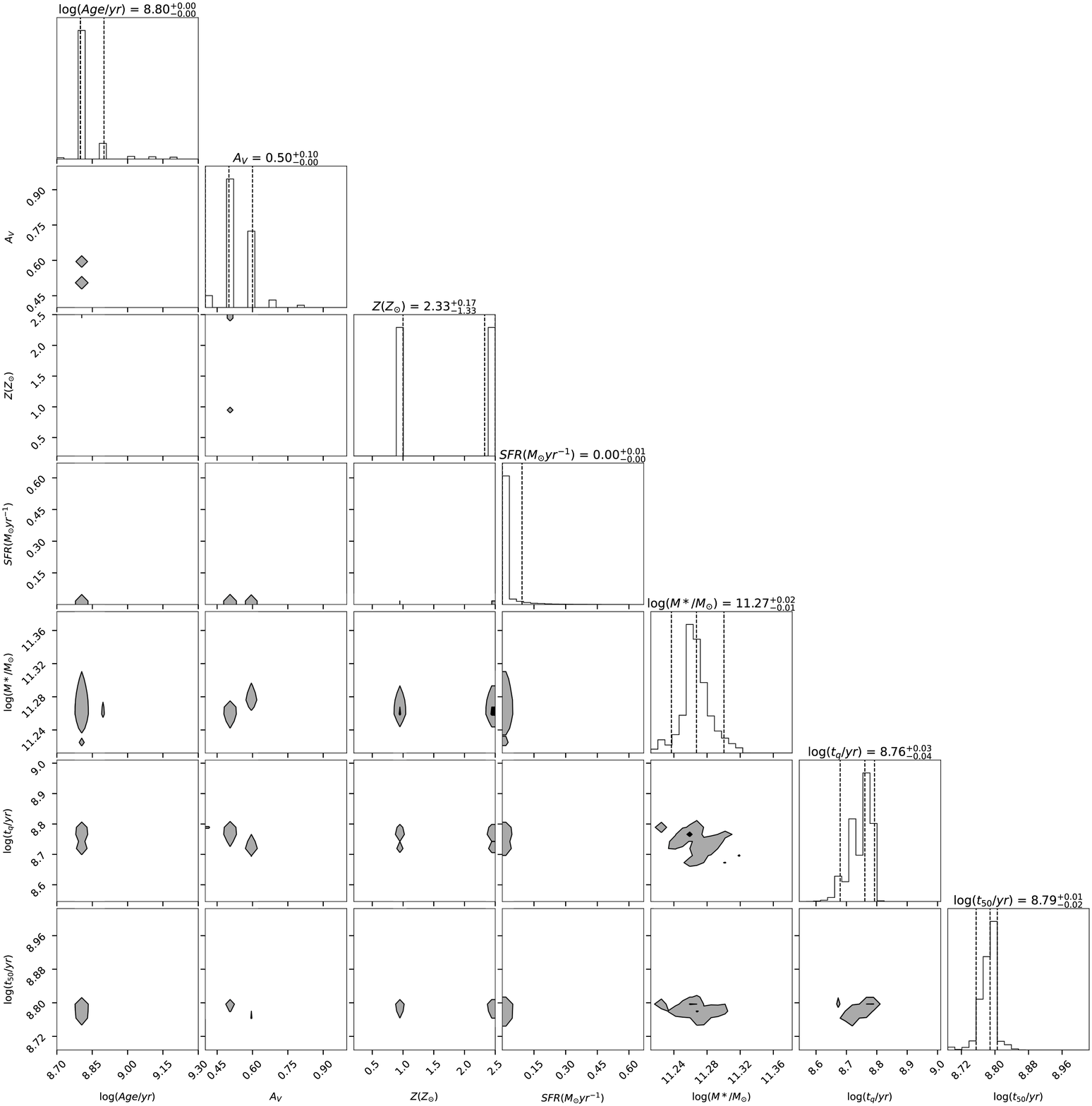}{1.0\textwidth}{}}
\caption{Corner plot for the ADF22-QG1 showing the SED fitting covariance 
between the age, $A_{\rm V}$, Z, SFR, $t_{\rm 50}$ and $t_{q}$.
The black and gray shaded region show the 68\% and 95\% confidence interval, respectively. 
The vertical dashed lines in the histograms show the 0.05, 0.5, and 0.95-quantiles of the probability distributions. 
\label{fig:cornerplot}}
\end{figure*}

Here we present the SFHs obtained with various models. 
First, we show the SFH and  the corner plot for the main quantities 
evaluated with {\sf FAST++} in Fig. \ref{fig:sfhfastpp} and Figure \ref{fig:cornerplot}, respectively.
The corner plots presented in this paper are generated with {\sf corner.py} \citep{2016JOSS....1...24F}.
In this model SFH, star formation occurs almost instantaneously. 
This model can explain the observed SED but results in unrealistically high maximum SFR. 

\begin{figure}
\gridline{\fig{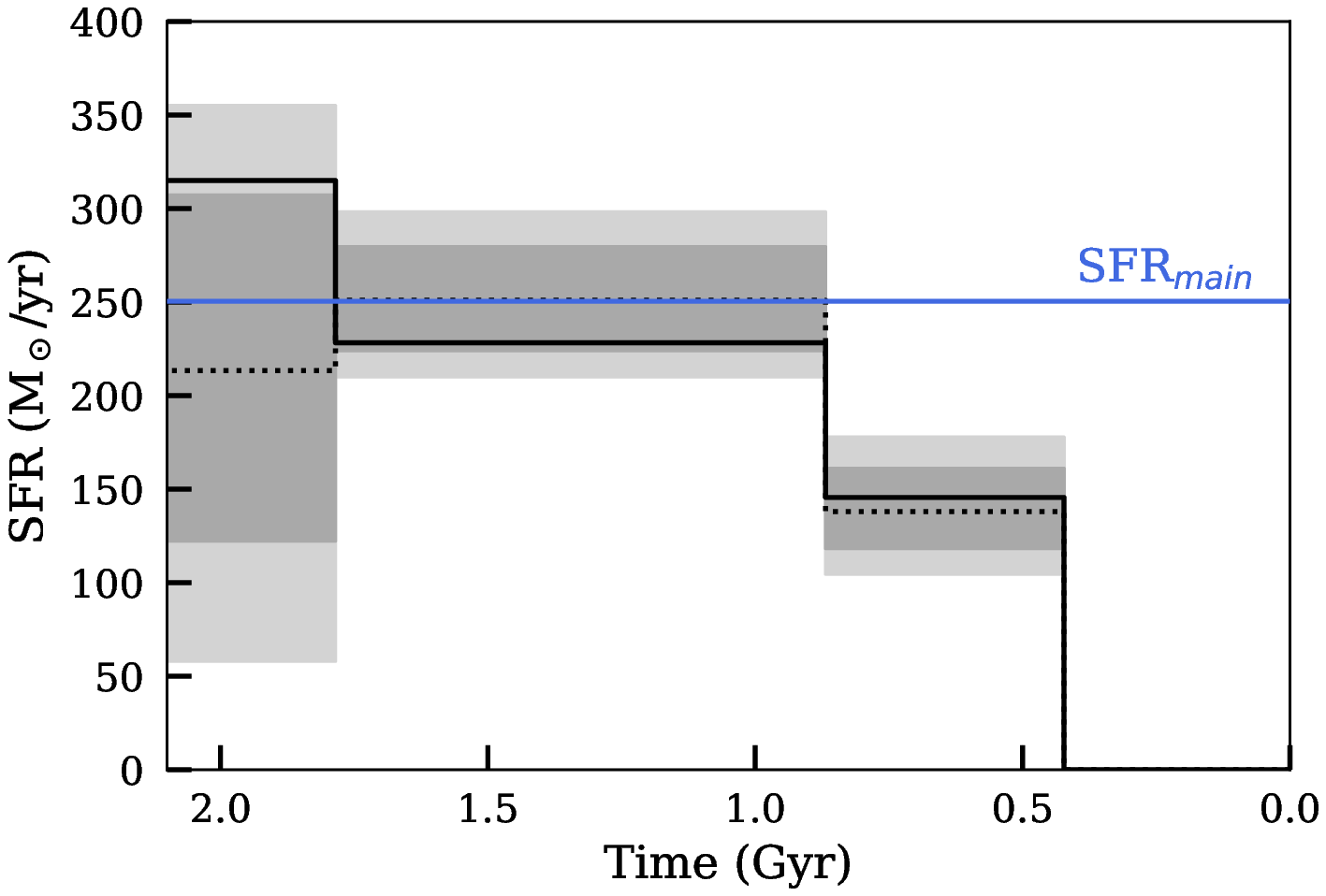}{0.5\textwidth}{}}
\gridline{\fig{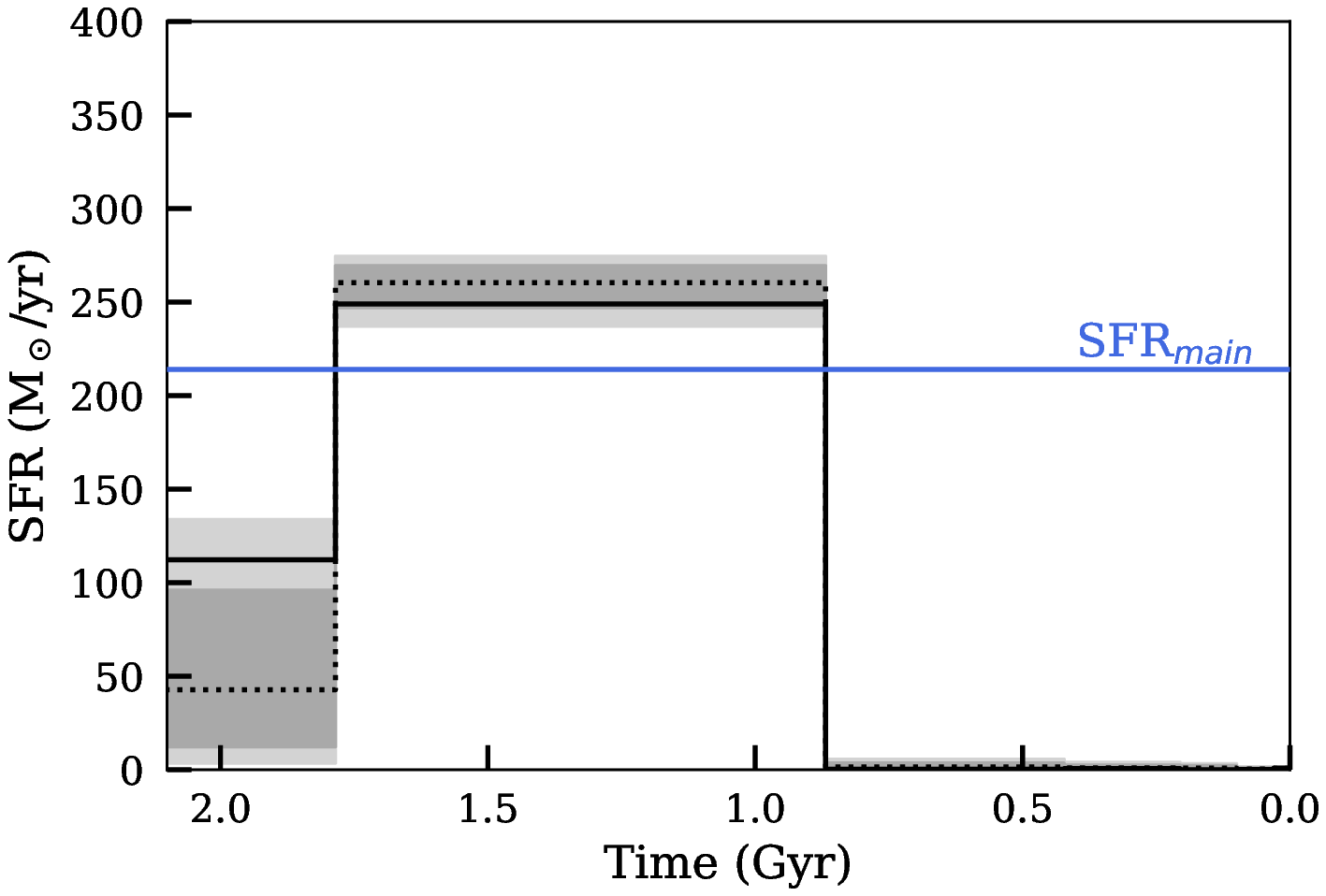}{0.5\textwidth}{}}
\gridline{\fig{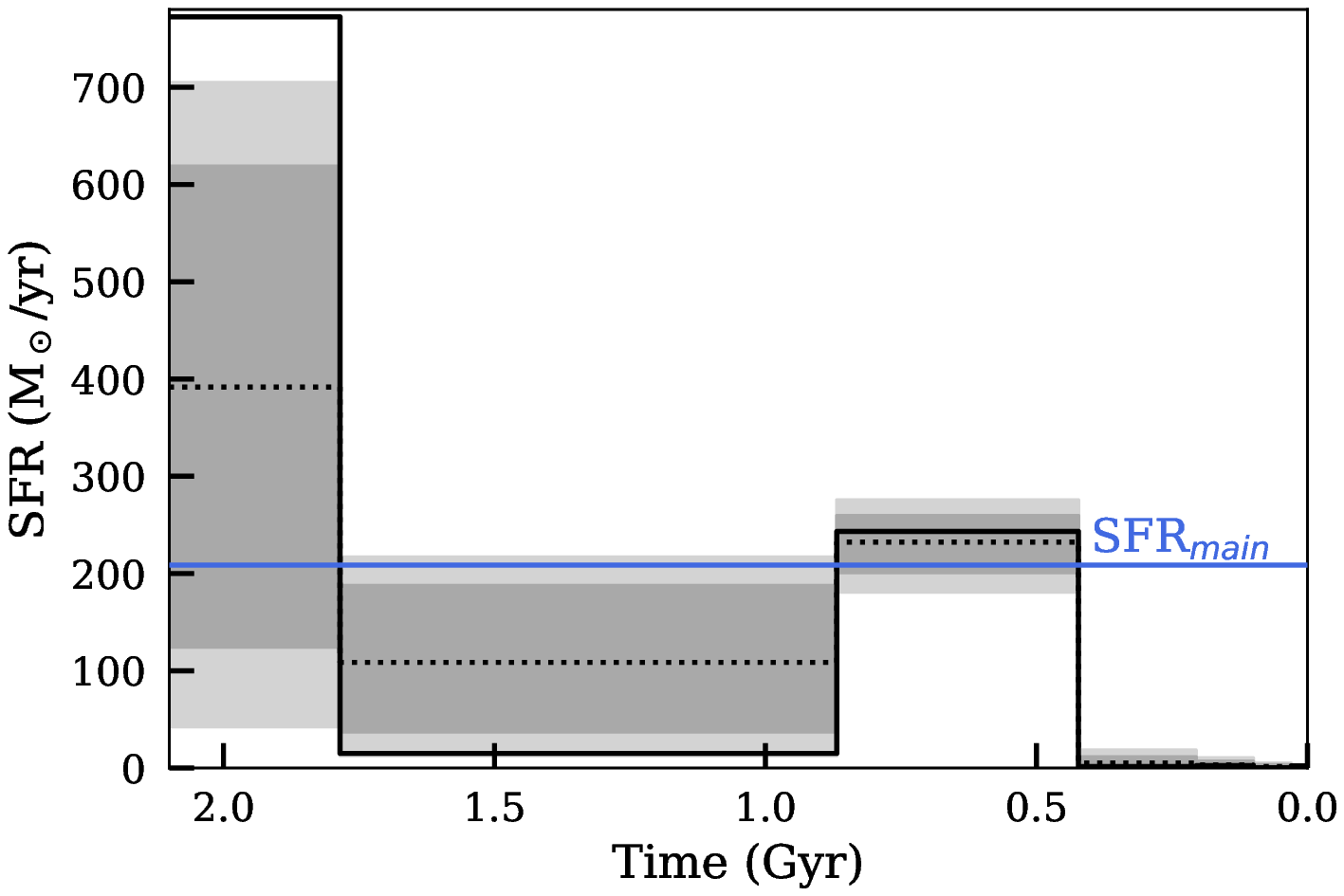}{0.5\textwidth}{}}
\caption{Similar to Fig. \ref{fig:sfh} but measured with nonparametric SFH modelings. 
They are modeled with ii) continuity SFH prior with free metallicity, 
iii) {\sf Prospector}-$\alpha$ SFH prior fixed with solar metallicity, 
and iv) {\sf Prospector}-$\alpha$  SFH prior with free metallicity, from {\it top} to {\it bottom}. 
\label{fig:sfhpros}}
\end{figure}

We also present the nonparametric SFHs evaluated with {\sf Prospector}. 
Here we show the SFHs evaluated adopting the models, 
i) Continuity SFH fixed with the solar metallicity (shown in Fig. \ref{fig:sfh} in main text), 
ii) Continuity SFH with free metallicity (-2.0 $\log Z/Z_{\odot}<0.2$), 
iii) {\sf Prospector}-$\alpha$ SFH fixed with the solar metallicity, and 
iv)  {\sf Prospector}-$\alpha$ SFH with free metallicity (-2.0 $\log Z/Z_{\odot}<0.5$) (shown in Fig. \ref{fig:sfhpros}).

The continuity SFH adopted in  i) and ii) models is described in section 2.5.1. 
The {\sf Prospector}-$\alpha$ model is the model adopted in \citet{2017ApJ...837..170L}. 
This model uses Dirichlet SFH prior, dust attenuation law prescription in \citet{2009A&A...507.1793N} and adds AGN component. 
In this prior, the fractional sSFR in each time bin follows a Dirichlet distribution. 
The dust attenuation model in {\sf Prospector} is \citet{2000ApJ...539..718C} model
consisting from birth-clouds and diffuse dust screen components.
Adopting the prescription by \citet{2009A&A...507.1793N}, 
the optical depth of the former follows 
$$ \hat{\tau}_{\lambda,1} =  \hat{\tau}_1 (\rm \lambda/5500\AA)^{-1.0},$$
and the latter follows 
$$ \hat{\tau}_{\lambda,2} =  \frac{\hat{\tau}_2}{4.05} (k'( \lambda) + D(\lambda))\left( \frac{\lambda}{\lambda_{V}}\right)^n,$$
where $\hat{\tau}_2$ controls the normalization of the diffuse dust, 
$n$ is the diffuse dust attenuation index,
$k'( \lambda)$ is the \citet{2001PASP..113.1449C} attenuation curve, 
and $D(\lambda)$ is a Lorentzian-like Drude profile describing the UV dust bump. 
The strength of the UV dust bump is tied with the best-fit diffuse dust attenuation index following the results of \citet{2013ApJ...775L..16K}.
A flat prior over $0<\hat{\tau}_{\lambda,2}<4$, a flat prior over $0<\hat{\tau}_{\lambda,1}<4$, 
and a flat prior over $-2.2<n<0.4$ are adopted. 
This flexible dust attenuation law can handle dust attenuation law variation 
where \citet{2013ApJ...775L..16K} reported that galaxies with lower specific SFR tend to have larger dust attenuation indices. 
Here the AGN SED templates in \citet{2008ApJ...685..147N,2008ApJ...685..160N} are adopted. 
The contribution of an AGN is controlled with the AGN luminosity as a fraction of the galaxy bolometric luminosity (f$_{\rm AGN}$) 
and optical depth of AGN torus dust ($\tau_{\rm AGN}$). 

\begin{figure*}
\gridline{\fig{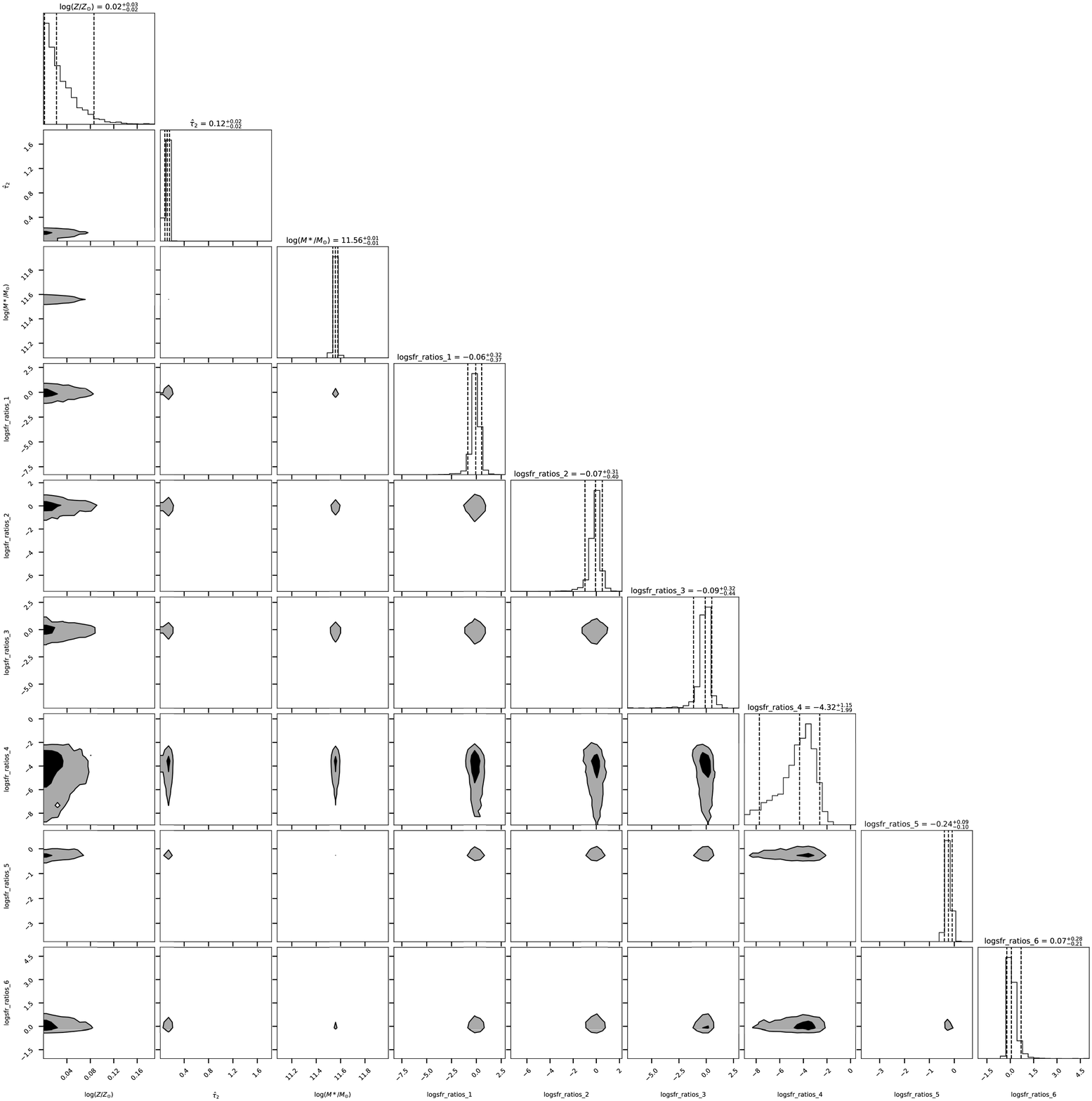}{1.0\textwidth}{}}
\caption{Similar to Fig. \ref{fig:cornerplot} but computed with {\sf Prospector} adopting the continuity SFH prior. 
The parameters are $\log Z/Z_{\odot}$, $\hat{\tau}_2$, $\log (M_{\star}/M_{\odot})$ (logmass),
and $\Delta\log$(SFR) between adjacent time bins (logsfr\_ratio\_n where $n=1-7$). 
\label{fig:cornerplot_continuity}}
\end{figure*}

\begin{figure*}
\gridline{\fig{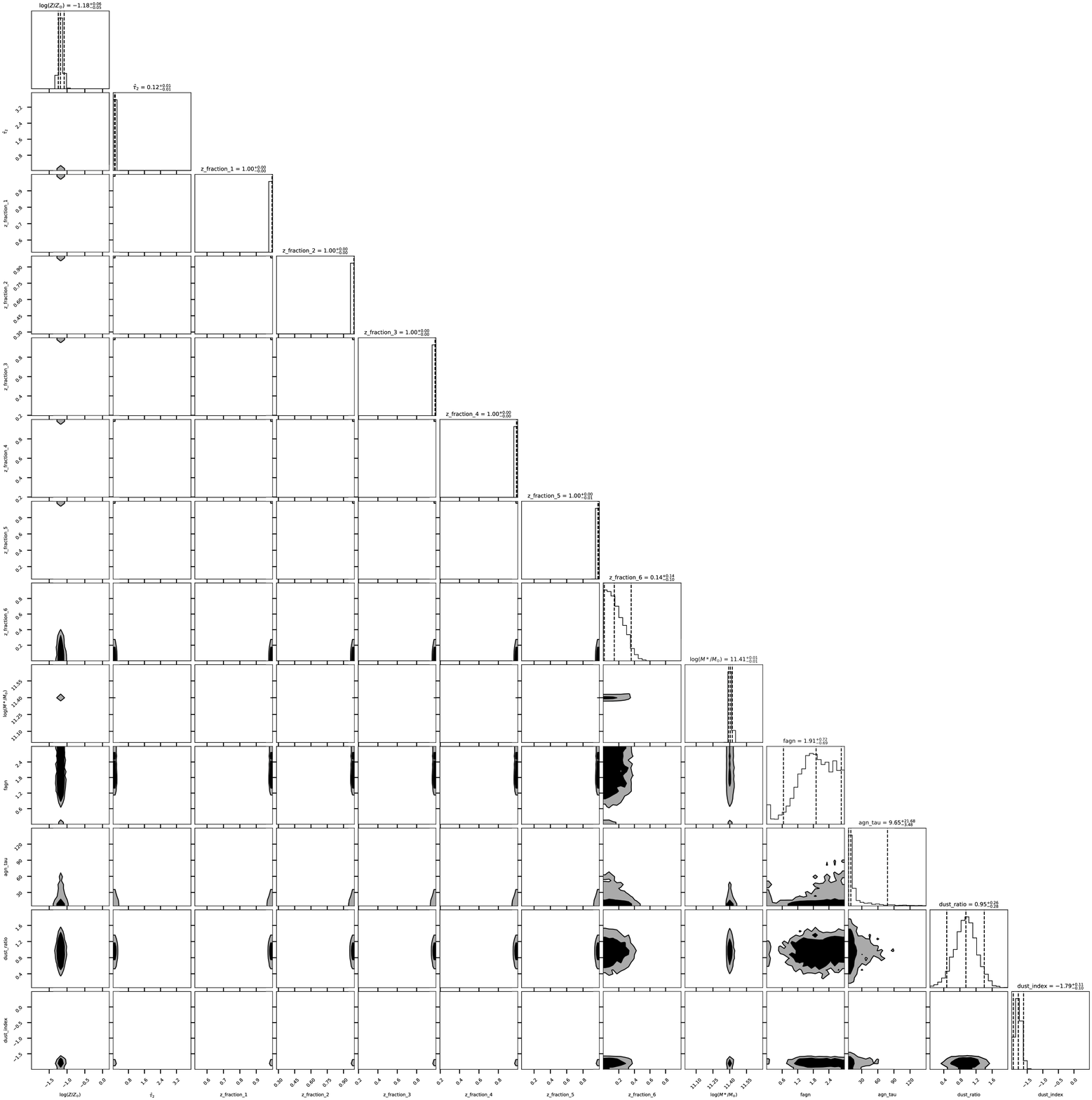}{1.0\textwidth}{}}
\caption{Similar to Fig. \ref{fig:cornerplot_continuity} but adopting the {\sf Prospector}-$\alpha$ SFH prior. 
The parameters are $\log Z/Z_{\odot}$, $\hat{\tau}_2$,  
z\_fraction which is proportional to sSFR between adjacent time bins (z\_fraction with $n=1-7$), 
$\log (M_{\star}/M_{\odot})$, AGN fraction (fagn), the optical depth for AGN (agn\_tau), 
the ratio of $\hat{\tau}_1$ and $\hat{\tau}_2$ (dust\_ratio),
and $n$ (dust\_index ). 
\label{fig:cornerplot_alpha}}
\end{figure*}

We show the SFH measured with ii) to iv) models in Fig. \ref{fig:sfhpros}.
The corner plots of the parameters for ii) and iv) models are shown in 
Fig. \ref{fig:cornerplot_continuity} and Fig. \ref{fig:cornerplot_alpha}, respectively. 
For the illustration purpose, the corner plots are shown for the modeling with the number of time bins $N=7$
which give consistent parameters as those evaluated with 14 time bins. 
We also test \citet{2001PASP..113.1449C} dust attenuation curve for Dirichlet prior 
but it does not change these parameters significantly. 
The AGN component in our target is negligible. 
From these above, the difference between the continuity SFH model and {\sf Prospector}-$\alpha$ model
mostly comes from the different SFH priors. 

All i) to iv) models results in similar $\hat{\tau}_2$ and stellar mass. 
They are in agreement with the quiescence of the star formation in our target.
In case of the continuity SFH prior, the modeling with free metallicity results in solar metallicity 
while that for {\sf Prospector}-$\alpha$ results in $\log Z/Z_{\odot}\sim-1$. 
As a result of the low luminosity, the SFH from iv) modeling quenched star formation earlier than those from other models. 
Since our observational data has no significant indices to confirm the metallicity robustly, 
we adopt the parameters measured adopting the continuity SFH prior and solar metallicity as conservative estimates. 

\counterwithin{figure}{section}

\bibliographystyle{aasjournal}



\end{document}